\documentclass[useAMS,usenatbib,pdftex]{mn2e}
\usepackage{float}
\usepackage{graphicx}
\usepackage{amsmath,amssymb}
\usepackage[dvips,usenames]{color}
\usepackage{hyperref}
\usepackage{tabularx}
\def\aj{AJ}%
\def\actaa{Acta Astron.}%
\def\apj{ApJ}%
\def\apjl{ApJ}%
\def\apjs{ApJS}%
\def\aap{A\&A}%
\def\aapr{A\&A~Rev.}%
\def\mnras{MNRAS}%
\def\pasp{PASP}%
\def\solphys{Sol.~Phys.}%
\newcommand{\sqdeg}{\mbox{deg$^{2}$}}

\newcommand{\jk}{\mbox{$J\!-\!K$}}
\newcommand{\yk}{\mbox{$Y\!-\!K$}}
\newcommand{\jks}{\mbox{$J\!-\!K_{\rm s}$}}
\newcommand{\yks}{\mbox{$Y\!-\!K_{\rm s}$}}
\newcommand{\ks}{\mbox{$K_{\rm s}$}}

\newcommand{\dmo}{\mbox{$(m\!-\!M)_{0}$}}

\newcommand{\av}{\mbox{$A_V$}}
\newcommand{\evi}{\mbox{$E_{V\!-\!I}$}}

\newcommand{\feh}{\mbox{\rm [{\rm Fe}/{\rm H}]}}
\newcommand{\mh}{\mbox{\rm [{\rm Fe}/{\rm H}]}}
\newcommand{\Msun}{\mbox{$M_{\odot}$}}

\newcommand{\logt}{\mbox{$\log (t/{\rm yr})$}}

\newcommand{\chisqmin}{\mbox{$\chi^2_{\rm min}$}}

\newcommand{\comment}[1]{}
\newcommand{\beq}{\begin{equation}}
\newcommand{\eeq}{\end{equation}}
\newcommand{\beqa}{\begin{eqnarray}}
\newcommand{\eeqa}{\end{eqnarray}}

\title[The SMC's SFH from VMC]{The VMC Survey -- XIV. First results on the 
look-back time star-formation rate tomography of the Small Magellanic Cloud\thanks{Based on observations made with VISTA at ESO under programme ID 179.B-2003}}
\author[Rubele et al.]{
Stefano Rubele$^{1}$, 
L\'eo Girardi$^{1}$, 
Leandro Kerber$^{2}$, 
Maria-Rosa L. Cioni$^{3,4}$, \newauthor 
Andr\'es E. Piatti$^{5,6}$, 
Simone Zaggia$^{1}$, 
Kenji Bekki$^{7}$,
Alessandro Bressan$^{8}$, \newauthor 
Gisella Clementini$^{9}$,
Richard de Grijs$^{10,11}$,
Jim P. Emerson$^{12}$, \newauthor 
Martin A.T. Groenewegen$^{13}$,  
Valentin D. Ivanov$^{14}$, 
Marcella Marconi$^{15}$, \newauthor
Paola Marigo$^{16}$,
Maria-Ida Moretti$^{9,17}$, 
Vincenzo Ripepi$^{15}$,
Smitha Subramanian$^{18}$,  \newauthor
Benjamin L. Tatton$^{19}$,
Jacco Th. van Loon$^{19}$
  \\
  $^1$ Osservatorio Astronomico di Padova -- INAF, Vicolo dell'Osservatorio 5, I-35122 Padova, Italy \\
  $^2$ Universidade Estadual de Santa Cruz, Rodovia Ilh\'eus-Itabuna, km 16, 45662-200 Ilh\'eus, Bahia, Brazil \\
  $^3$ University of Hertfordshire, Physics Astronomy and Mathematics, College Lane, Hatfield AL10 9AB, UK \\
  $^4$ Leibnitz-Institut f\"ur Astrophysik Potsdam, An der Sternwarte 16, 14482 Potsdam, Germany\\
  $^5$ Observatorio Astron\'omico, Universidad Nacional de C\'ordoba, Laprida 854, 5000, C\'ordoba, Argentina\\
  $^6$ Consejo Nacional de Investigaciones Cient\'{\i}ficas y T\'ecnicas, Av.\ Rivadavia 1917, C1033AAJ, Buenos Aires, Argentina \\
  $^7$ ICRAR, M468, The University of Western Australia, 35 Stirling Hwy, Crawley, WA 6009, Australia \\
  $^8$ SISSA, via Bonomea 265, I-34136 Trieste, Italy \\
  $^9$ INAF-Osservatorio Astronomico di Bologna, via Ranzani 1, I-40127 Bologna, Italy \\
  $^{10}$ Kavli Institute for Astronomy \& Astrophysics, Peking University, Yi He Yuan Lu 5, Hai Dian District, Beijing 100871, China \\
  $^{11}$ Department of Astronomy, Peking University, Yi He Yuan Lu 5, Hai Dian District, Beijing 100871, China \\
  $^{12}$ Astronomy Unit, School of Physics and Astronomy, Queen Mary University of London, Mile End Road, London E1 4NS, UK \\
  $^{13}$ Koninklijke Sterrenwacht van Belgi\"e, Ringlaan 3, B--1180 Brussels, Belgium \\
  $^{14}$ ESO European Southern Observatory, Ave. Alonso de Cordova 3107, Casilla 19, Chile\\
  $^{15}$ INAF-Osservatorio Astronomico di Capodimonte, Via Moiariello 16, 80131, Naples, Italy \\
  $^{16}$ Dipartimento di Fisica e Astronomia, Universit\`a di Padova, Vicolo dell’Osservatorio 2, I-35122 Padova, Italy \\
  $^{17}$ Scuola Normale Superiore, Piazza dei Cavalieri 7, I-56126 Pisa, Italy \\
  $^{18}$ Indian Institute of Astrophysics, Koramangala II Block, 560034 Bangalore, India \\
  $^{19}$ Lennard-Jones Laboratories, Keele University, ST5 5BG, UK
}

\begin{document}

\date{ MNRAS accepted
}

\pagerange{\pageref{firstpage}--\pageref{lastpage}} \pubyear{2013}

\maketitle

\label{firstpage}

\begin{abstract}
We analyse deep images from the VISTA survey of the Magellanic Clouds in the $YJ\ks$ filters, covering 14~\sqdeg\ (10 tiles), split into 120 subregions, and comprising the main body and Wing of the Small Magellanic Cloud (SMC). We apply a colour--magnitude diagram reconstruction method that returns their best-fitting star formation rate SFR$(t)$, age--metallicity relation (AMR), distance and mean reddening, together with 68\% confidence intervals. The distance data can be approximated by a plane tilted in the East--West direction with a mean inclination of $39^\circ$, although deviations of up to $\pm3$~kpc suggest a distorted and warped disk. 
After assigning to every observed star a probability of belonging to a given age--metallicity interval, we build high-resolution population maps. These dramatically reveal the flocculent nature of the young star-forming regions and the nearly smooth features traced by older stellar generations. They document the formation of the SMC Wing at ages $<\!0.2$~Gyr and the peak of star formation in the SMC Bar at $\sim\!40$~Myr. We clearly detect periods of enhanced star formation at 1.5 Gyr and 5 Gyr. The former is possibly related to a new feature found in the AMR, which suggests ingestion of metal-poor gas at ages slightly larger than 1 Gyr. The latter constitutes a major period of stellar mass formation. We confirm that the SFR$(t)$ was moderately low at even older ages.
\end{abstract}


\section{Introduction}
\label{sec:intro}

Owing to its proximity, significant mass, and ongoing star formation, the Small Magellanic Cloud (SMC) represents a fundamental laboratory for studies of stellar and interstellar medium (ISM) processes. It contains a rich population of star clusters, associations, stellar pulsators, primary distance indicators, and stars in short-lived evolutionary stages \citep{westerlund90, bica95, bolatto07, soszynski10b, soszynski10a, soszynski11}, with metallicities systematically lower than those observed in the Large Magellanic Cloud (LMC) and Milky Way (MW) galaxies. 

The SMC is probably the largest among the Local Group galaxies in which a disk cannot be identified at first sight. Its optical appearance on the sky is dominated by an elongated bar-like structure in the direction NE--SW, from which a prominent Wing departs to the east, in a direction that joins the Magellanic Bridge and the LMC \citep{nidever_etal11}. In addition, there have been consistent claims of depth structures being detected along different lines-of-sight \citep{gardiner91, nidever13}, and the growing awareness that the Bridge is, at least partially, a tidally-stripped component extending towards the LMC \citep{har07, nidever13}. Needless to say, any work on stellar or ISM physics can only be complicated by this particular morphology. 
 
The star formation history (SFH) of the SMC also carries along the signs of a disturbed past, probably marked by the interactions with the LMC\footnote{Although possible past interactions with the Milky Way are presently disfavoured \citep{besla07}, they might have occurred at ages larger than 4~Gyr, depending on the precise mass of the galaxies involved \citep{kal13, gomez14}.}. The presence of periods of enhanced star and cluster formation has for long been recognized, as well as a relative paucity of very old populations \citep[e.g.][]{HZ04, degrijs08, netal09, p11, cignoni12,cignoni13, wetal13}\footnote{The presence of a very old population is demonstrated by the presence of 2475 RR~Lyrae  \citep{soszynski10a} detected by the OGLE~III survey.}. This disturbed past became even more evident with the discovery that a fraction of SMC stars has been captured by the LMC \citep{Olsen_etal11}.

The VISTA Survey of the Magellanic Clouds \citep[VMC;][]{Cioni11} was designed to greatly enhance our knowledge of the Magellanic system, via near-infrared photometry of the bulk of their stars, down to depths reaching the oldest turn-offs in both the LMC and SMC \citep{Kerber_etal09}. The advantage over previous deep-and-wide surveys resides mainly in the use of near-infrared wavelengths, that provide a view of the stellar populations almost unaffected by dust. The near-infrared light is also dominated by cool giants and subgiants, hence better sampling the intermediate-age populations over the young ones. In addition to depth, VMC provides twelve-epochs photometry in the \ks-band, allowing us to derive tighter period--luminosity relations for variables and more stringent constraints on the distances \citep[e.g.][]{Ripepi_etal12} and possible structures along the line of sight \citep{Moretti14}.
 
In this paper, we analyse the VMC photometry of the SMC in terms of SFH and geometry. It constitutes the natural follow-up of the papers by \citet{Kerber_etal09} and \citet{Rubele_etal12}, in which the method for SFH-recovery from VMC data was first planned and tested, and the first analysis of the LMC data was performed, respectively. Since then, an impressive amount of data has been collected for the SMC galaxy, allowing us to perform a similar analysis. Present data for the SMC comprise most of its stellar mass, all of its bar, and a significant fraction of its Wing, as well as a few outer regions.

The structure of this paper is as follows. Sect.~\ref{sec:data} presents the new VMC data and the PSF photometry.  Sect.~\ref{sec:sfh} details the method adopted for the derivation of the SFH and AMR, and their errors, as well as the best-fitting mean distances and reddenings for every subregion of a tile. Sect.~\ref{sec:close} discusses all the derived trends, also in comparison with previous results in the literature. Finally Sect.~\ref{sec_summary} summarizes the main results and draws a few concluding comments.

\begin{table*}[t]
  \caption{VMC tiles used in this work.
  }
\label{tab_tiles}
\begin{tabular}{ccccccccc}
\hline
\hline
Tile & $\alpha$ (deg, J2000) & $\delta$ (deg, J2000) & Completion in \ks-band$^1$ & Comments\\
\hline
SMC 3\_3 & 11.1995 & $-$74.2005 & 92\% & S extreme of bar\\
SMC 3\_5 & 21.9762 & $-$74.0018 & 100\% & S part of Wing\\
SMC 4\_3 & 11.2810 & $-$73.1094 & 88\% & SW part of densest bar \\
SMC 4\_4 & 16.3303 & $-$73.0876 & 92\% & central bar, slightly towards Wing\\
SMC 4\_5 & 21.2959 & $-$72.9339 & 92\% & central part of Wing\\
SMC 5\_3 & 11.2043 & $-$72.0267 & 92\% & NW of densest bar\\
SMC 5\_4 & 16.1918 & $-$71.9850 & 100\% & NE part of densest bar\\ 
SMC 5\_6 & 25.4401 & $-$71.5879 & 88\% & $\sim4$~deg E of main body \\
SMC 6\_3 & 11.4311 & $-$70.9266 & 78\% & $\sim1.5$~deg NW of main body\\
SMC 6\_5 & 20.4138 & $-$70.7601 & 78\% & $\sim3$~deg NE of main body\\
\hline
\end{tabular}

$^1$ Here $100$\% completion corresponds to $12$ epochs in the \ks-band or at least $9000$ sec of integration time.
\end{table*}

\begin{figure*}
\resizebox{0.8\hsize}{!}{\includegraphics{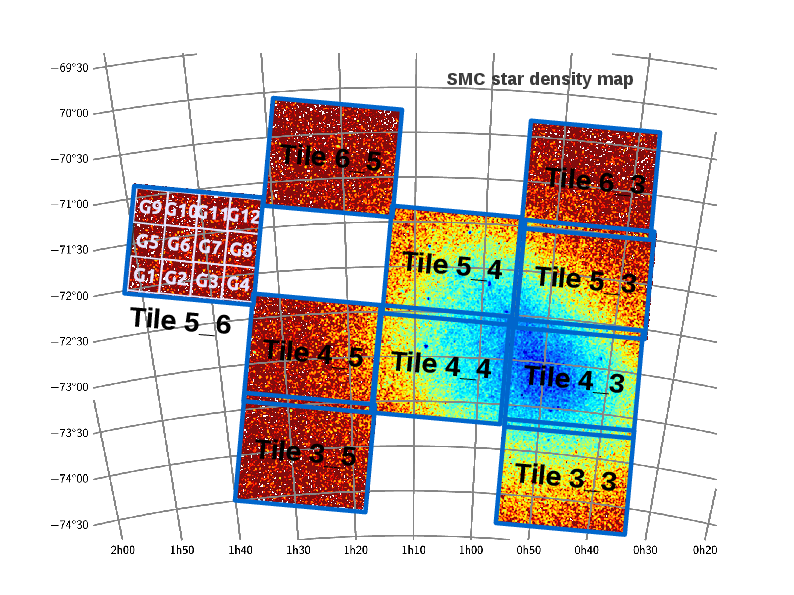}}
\caption{Sky position of all tiles analysed in this work, as listed in Table~\ref{tab_tiles}. Each tile is subdivided into 12 small subregions, as illustrated here for the tile 5\_6. The plot includes only the tiles with a high fraction of the planned VMC observations already completed; the complete, final coverage of the SMC by the VMC survey can be appreciated in \citet{Cioni11}.}
\label{fig_obstiles}
\end{figure*}

\section{Data and photometry}
\label{sec:data}

\subsection{Selected VMC data}
\label{sec:selecteddata}
  
All data used in this work come from the VMC survey, which is extensively described in \citet{Cioni11}.
We use the v1.1 and v1.2 VMC data retrieved from the VISTA Science Archive \citep[VSA;][]{Hambly_etal04}\footnote{\url{http://horus.roe.ac.uk/vsa/}}. More specifically, we start from the pawprints already processed by the VISTA Data Flow System \citep[VDFS;][]{Emerson_etal04} in its pipeline \citep{Irwin_etal04}. Individual pawprints are combined into deep tiles as described in Appendix~\ref{sec:vmcdata}. The photometric zeropoints are discussed in Appendix~\ref{sec:zeropoints}. 

In this work we deal with the the ten SMC tiles for which we already have collected a large fraction/the whole set of \ks\ multi-epoch photometry. They are illustrated in Fig.~\ref{fig_obstiles} and listed in Table~\ref{tab_tiles}. 
The background image in Fig.~\ref{fig_obstiles} is a density map of all VMC sources with $\ks\!<\!18$~mag and $\ks$ errors smaller than $0.2$~mag. Since this magnitude cut includes the red clump (RC) and the upper part of the red giant branch (RGB), the map is dominated by the intermediate-age and old stellar populations. It is immediately evident that our ten tiles cover most of the SMC main body, including all of its dense bar, which appears in Fig.~\ref{fig_obstiles} as a bean-like blue region across the tiles SMC 3\_3, 4\_3, 4\_4, 5\_3 and 5\_4. In addition, the region analysed includes a large fraction of the so-called Wing (tiles SMC 3\_5, 4\_5 and 5\_6), and more external regions towards the N and NE of the SMC (tiles SMC 6\_3 and 6\_5). 
We can confidently state that we have observed the majority of the SMC stellar mass, and the bulk of the young star formation. As we will see later, these observations also span a wide variety of SMC environments and distances.

\subsection{PSF photometry and artificial star tests}
\label{sec:psfphotometry}

We have produced homogenised deep tile images and performed PSF photometry on them, as described in Appendix~\ref{sec:vmcdata}. Afterwards we correlated the three bands ($YJ\ks$) photometry using a $1\arcsec$ matching radius generating a multi-band catalog. Finally, we corrected our photometry for the aperture using as reference the VSA data release v1.2 (see \citealt{Cross_etal12} and \citealt{Irwin_etal04} for details).

\begin{figure*}
\resizebox{0.24\hsize}{!}{\includegraphics{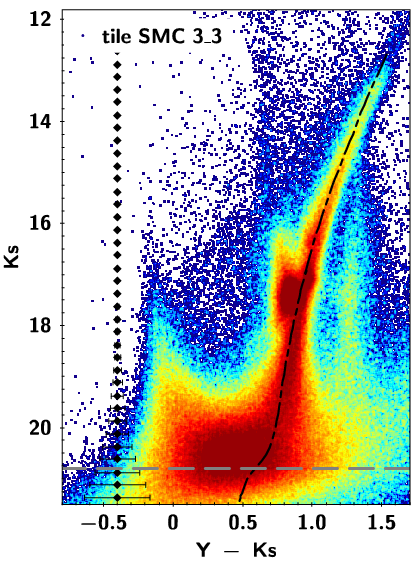}}
\resizebox{0.24\hsize}{!}{\includegraphics{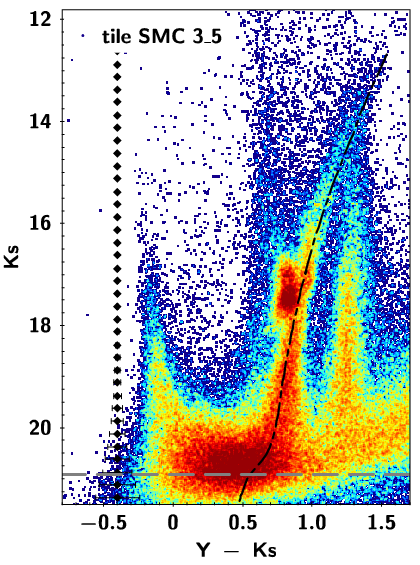}}
\resizebox{0.24\hsize}{!}{\includegraphics{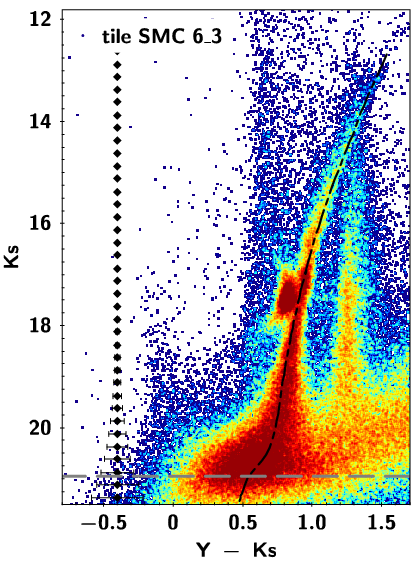}}
\resizebox{0.24\hsize}{!}{\includegraphics{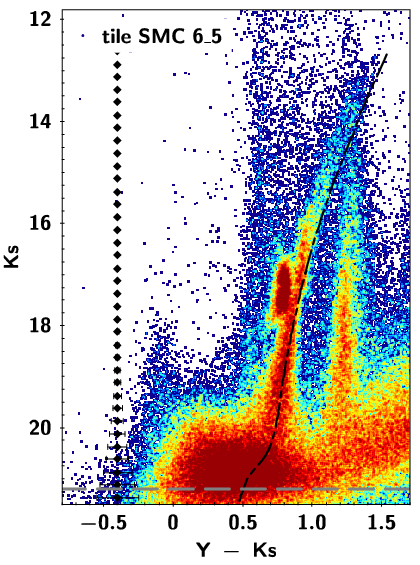}}
\\
\resizebox{0.24\hsize}{!}{\includegraphics{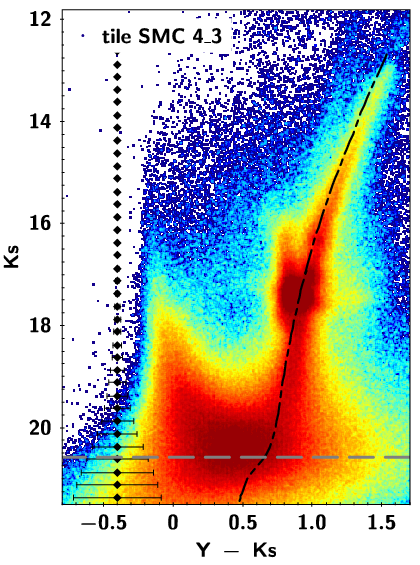}}
\resizebox{0.24\hsize}{!}{\includegraphics{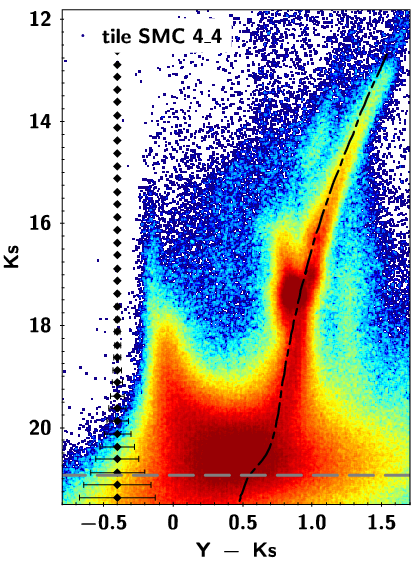}}
\resizebox{0.24\hsize}{!}{\includegraphics{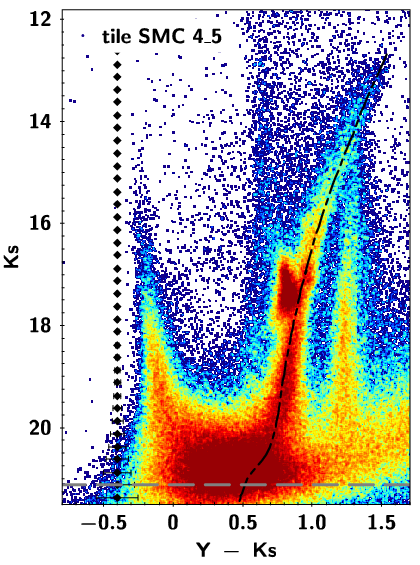}}
\\
\resizebox{0.24\hsize}{!}{\includegraphics{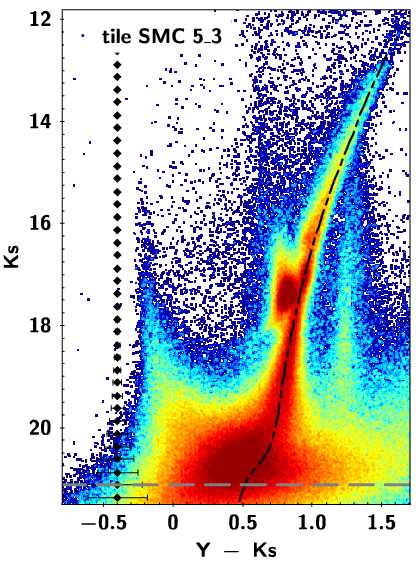}}
\resizebox{0.24\hsize}{!}{\includegraphics{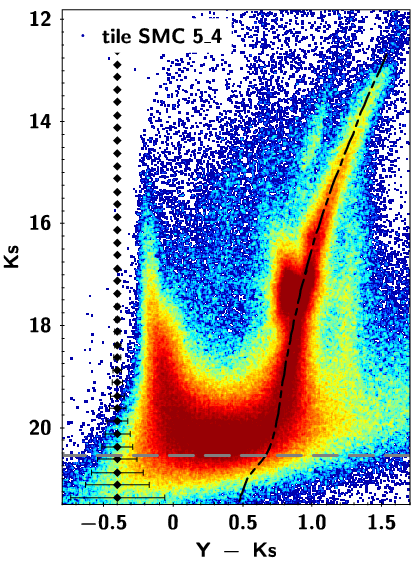}}
\resizebox{0.24\hsize}{!}{\includegraphics{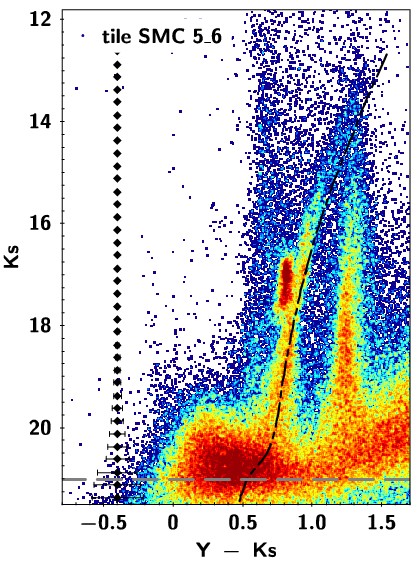}}
\caption{PSF photometry for the ten tiles analysed in this work, in the $\ks$ versus $Y\!-\!\ks$ diagram. Colours from dark blue to red code the increasing density of stars in the CMD while the black bars on the left-hand side show the $\pm1\sigma$ errors in $\ks$-band. The grey dashed lines mark the 50~\% completeness level in the \ks\ band. Finally, the dark dot-dashed lines show the main sequence to RGB section of a 10~Gyr old isochrone of metallicity $\mh=-1$, shifted by a distance modulus of $\dmo=18.92$~mag and an extinction of $\av=0.35$~mag, for comparison.}
\label{YKcmd}
\end{figure*}

Figure \ref{YKcmd} shows the CMDs in $Y\!-\!\ks$ vs.\ $\ks$ for the ten SMC tiles investigated to date. The different colours map the density of stars in the CMD. These CMDs will be further discussed in the next sections, together with the derived SFHs.

Very extensive artificial star tests (ASTs) were performed on the homogenized images, so as to completely map the distributions of photometric errors and completeness, as a function of colour, magnitude, and position. The process is the same as extensively described and illustrated in \citet{Rubele_etal12}, with the only difference being that the photometry was made on the entire tile
(see Appendix~\ref{sec:vmcdata}). Suffice it to mention here that our typical 50\% completeness limit in the outermost tiles turns out to be at about 22.3, 22.1 and 20.8 mag in the $Y$, $J$ and $\ks$ filters, respectively. In the more central tile SMC 4\_3, the same limits are found at about 21.3, 21.1 and 20.6 mag, respectively.

In the following, we will only use VMC data and CMDs at magnitudes brighter than $\ks<20.5$~mag, hence ensuring that completeness never falls below the 50\% level.

\subsection{Defining subregions}

All tiles, covering total areas of $\sim 1.5$~\sqdeg\ each, were subdivided into twelve subregions of $21.0\arcmin\times21.5\arcmin$ ($\sim 0.12$~\sqdeg) for the SFH analysis. They are numbered from G1 to G12 as already illustrated in Fig.~\ref{fig_obstiles}. Central coordinates are listed in Table~\ref{tab_D_Av} below. G1 is located at the SE extreme of each tile, G4 at the SW, G9 at the NE, and G12 at the NW. 

The pawprints contributing to the corner subregion G1 (see Fig.~\ref{fig_obstiles}) include a contribution from the ``top'' half of the VIRCAM detector number 16, which is known to show a significantly worse signal-to-noise ratio than the other detectors \citep[see][for details]{Rubele_etal12}. The effect is negligible in \ks, small in $H$ and becomes more noticeable in the bluest VISTA passbands, i.e.\ in $J$, $Y$ and $Z$. Therefore part of subregions G1 of all tiles were excluded from our analysis.

\section{Deriving the SFH}
\label{sec:sfh}

\subsection{Overview of the method}
\label{sec:method}

\begin{figure}
\resizebox{0.32\hsize}{!}{\includegraphics{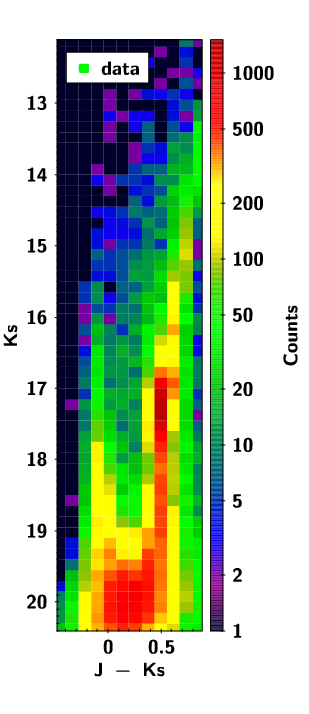}}
\resizebox{0.32\hsize}{!}{\includegraphics{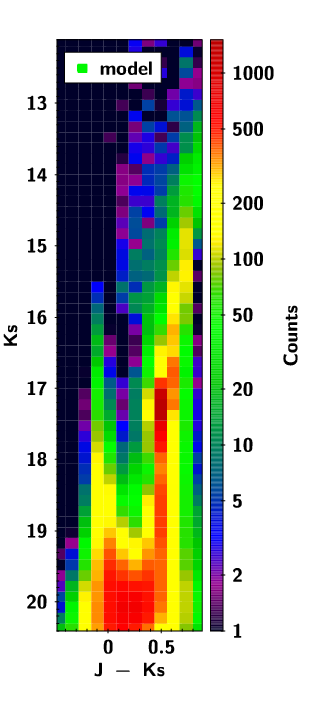}}
\resizebox{0.32\hsize}{!}{\includegraphics{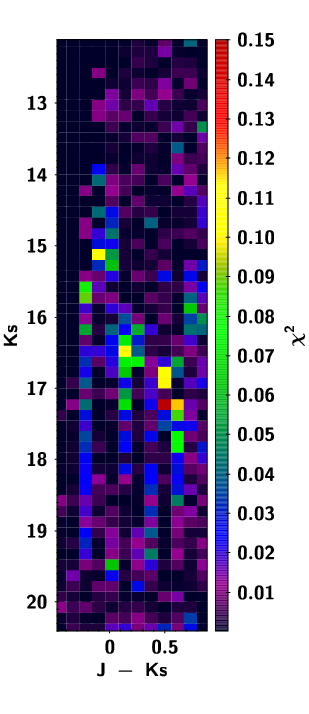}}
\\
\resizebox{0.32\hsize}{!}{\includegraphics{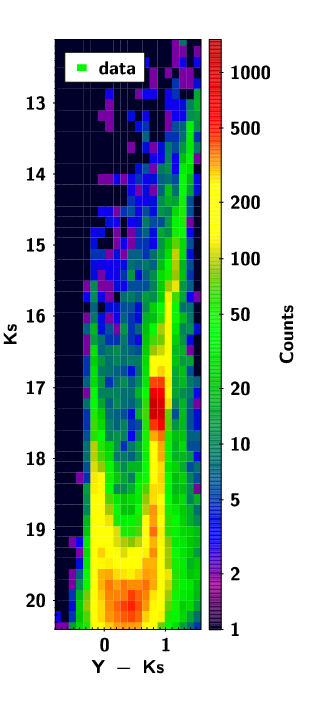}}
\resizebox{0.32\hsize}{!}{\includegraphics{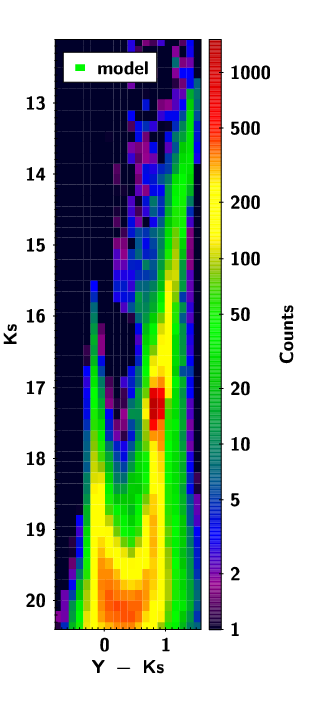}}
\resizebox{0.32\hsize}{!}{\includegraphics{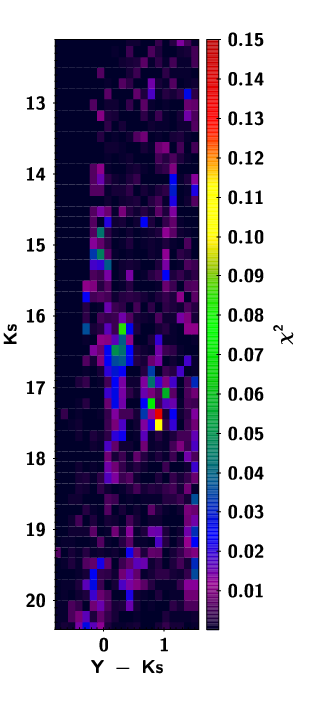}}
\caption{From left to right: Hess diagrams for the data, best fitting model, and chi-square, for the subregion G8 of tile SMC 5\_4. The top panels are for the \jks\ vs.\ks\ CMD, the bottom ones for \yks\ vs. \ks\ CMD. In the left and central panels, the colour scale indicates the number of stars per bin.}
\label{fig_sol}
\end{figure}

The method for deriving the SFH from VMC data has been already described extensively in \citet{Rubele_etal12}. As a short summary, we recall that, for every subregion of a tile:
\begin{enumerate}
\item We start assuming an absolute distance modulus, $\dmo$ and $V$-band extinction, $A_V$. 
\item ``Partial models'' are derived for the entire range of metallicities and ages of relevance (see next section). Partial models are simply synthetic stellar populations each covering a small bin of age and metallicity, shifted to the desired distance and extinction, and representing a given total mass of formed stars. The extinction in each VISTA passband is derived from the assumed $A_V$ using the extinction coefficients computed from the \citet{Cardelli_etal89} extinction curve with $R_V=3.1$ \citep[see][for details]{Girardi_etal08}, namely  $A_Y=0.385\av$, $A_J=0.283\av$ and $A_{K_{\rm s}}=0.114\av$\footnote{We choose to express the extinction in the $V$ band just for convenience in the subsequent comparisons with other authors (Sect.~\ref{sec:extinctionmap}), although we will be actually dealing with the extinction in near-infrared pass-bands. Notice that the average LMC and SMC extinction curves are essentially indistinguishable from the MW one for all wavelengths redward of $\sim\!5000$~\AA\ \citep[see e.g.][figure 10]{gordon03}, so that the choice of a near-infrared reference wavelength for the extinction, instead of $\av$, would lead essentially to the same results.} \label{item_alav}.
\item Finally, partial models are ``degraded'' to the conditions of the actual observations by applying the distributions of photometric errors and completeness implied by the ASTs, and translated into Hess diagrams, i.e.\ maps of the stellar density across the \ks\ versus $\jks$ and \ks\ versus $\yks$ CMDs. This involves a few steps: First, high-resolution Hess diagrams, with a resolution of 0.05~mag in magnitude and 0.04~mag in colour, are built from the synthetic stellar populations. For each one of these small boxes, all artificial stars of similar colour and magnitude are looked for in the AST database, in order to build two-dimensional histograms of the differences between the input and output magnitudes and colours, and compute the completeness fraction per box. The high-resolution Hess diagram is then degraded by applying these error distributions and completeness on a box-by-box basis, which has the effect of both spreading stars into the neigbouring boxes in the CMD, and reducing their numbers. Finally, the high-resolution degraded partial models are converted into Hess diagrams of normal resolution (namely 0.15~mag per 0.12~mag), which are used in the subsequent steps of the analysis.
\item The linear combination of partial models that best-fit the observed Hess diagrams is found by the StarFISH optimization code by \citet{HZ01}, suitably adapted to our case. An example of a best-fit model is presented in Fig.~\ref{fig_sol}. The coefficients of this linear combination of partial models are straightforwardly converted into the SFH. The best-fit solution is also characterized by the $\chisqmin$, that mesures the residuals between the best-fit model and the data.
\item The same process is repeated over a range of different $\dmo$ and $A_V$ values. This allows us to pin down the best-fit SFH, distance and extinction for each subregion, by simply looking for the minimum $\chisqmin$.
\item One hundred synthetic realizations of the best-fit model are generated and are analysed via the same method. The dispersion in the resulting coefficients provides us the confidence levels of our best-fit SFH, $\dmo$ and $A_V$.
\end{enumerate}
Several aspects of this process will be either better detailed or simply illustrated in the following, whenever relevant to understand the particular results we find for the SMC. 

\subsection{Partial models}
\label{sec:partialmodels}

\begin{table}
\caption{Grid of SMC stellar partial models used in the SFH recovery.}
\label{tab_amr}
\centering
\begin{tabular}{c|ccccc}
\hline\hline
$\log(t/{\rm yr})$ & $\feh_1$ & $\feh_2$ & $\feh_3$ & $\feh_4$ & $\feh_5$ \\
\hline
6.9 & $-$0.10 & $-$0.25 & $-$0.40 & $-$0.55 & $-$0.70 \\
7.4 & $-$0.10 & $-$0.25 & $-$0.40 & $-$0.55 & $-$0.70 \\
7.8 & $-$0.10 & $-$0.25 & $-$0.40 & $-$0.55 & $-$0.70 \\
8.1 & $-$0.10 & $-$0.25 & $-$0.40 & $-$0.55 & $-$0.70 \\
8.3 & $-$0.20 & $-$0.35 & $-$0.50 & $-$0.65 & $-$0.80 \\
8.5 & $-$0.20 & $-$0.35 & $-$0.50 & $-$0.65 & $-$0.80 \\
8.7 & $-$0.20 & $-$0.35 & $-$0.50 & $-$0.65 & $-$0.80 \\
8.9 & $-$0.40 & $-$0.55 & $-$0.70 & $-$0.85 & $-$1.00 \\
9.1 & $-$0.55 & $-$0.70 & $-$0.85 & $-$1.00 & $-$1.15 \\
9.3 & $-$0.55 & $-$0.70 & $-$0.85 & $-$1.00 & $-$1.15 \\
9.5 & $-$0.70 & $-$0.85 & $-$1.00 & $-$1.15 & $-$1.30 \\
9.7 & $-$0.85 & $-$1.00 & $-$1.15 & $-$1.30 & $-$1.45 \\
9.9 & $-$1.15 & $-$1.30 & $-$1.45 & $-$1.60 & $-$1.75 \\
10.075 & $-$1.45 & $-$1.60 & $-$1.75 & $-$1.90 & $-$2.05 \\
\hline
\end{tabular}
\end{table}

Partial models for this work have been derived from PARSEC v1.0 evolutionary tracks \citep{Bressan_etal12}, which represent a major revision and update of the previous sets of Padova evolutionary tracks \citep[namely ][]{Marigo_etal08, Girardi_etal10} used in \citet{Rubele_etal12}. The most relevant changes, in the context of this work, are
\begin{itemize}
\item the adoption of \citet[][and references therein]{Caffau_etal11} solar composition, with a present-day solar metal content of $Z=0.0152$;
\item the recalibration of the Solar Model so as to meet the stringent constraints imposed by helioseismology and by the radiometric age of the Solar System; this implies allowing microscopic diffusion to operate in low mass stars;
\item revised limits for the mass interval for which overshooting increases its efficiency, in low-mass stars.
\end{itemize}
The main consequence of these novelties are modest (though systematic) changes in the age and metallicity scale of stellar populations, at ages exceeding a few Gyr. The age scale is presently being tested on LMC/SMC intermediate-age star clusters for which we have HST data of excellent quality \citep[e.g.][]{Girardi_etal13, correnti14, goudfrooij14}, so that we postpone further comments on this topic to future papers. Also, we recall that the present release of PARSEC tracks (v1.1 as of this writing) already includes modifications in the prescriptions for microscopic diffusion in low-mass stars, with respect to v1.0. We have checked that these changes are simply negligible in the range of magnitudes and colours sampled by the VMC data for the SMC.
 
\begin{figure*}
\resizebox{0.54\hsize}{!}{\includegraphics{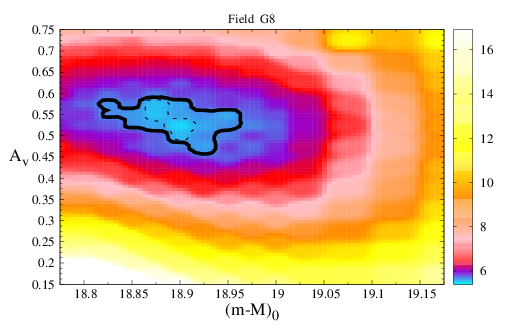}}\hfill
\resizebox{0.44\hsize}{!}{\includegraphics{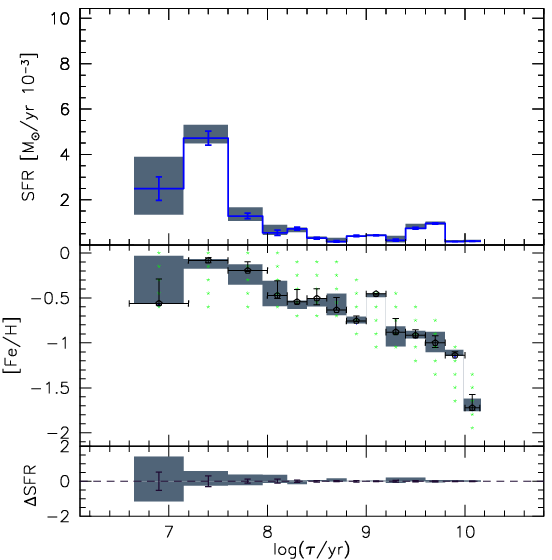}}
\caption{The SFH results for subregion G8 of tile SMC 5\_4, which corresponds to a relatively dense region of the SMC bar. The {\bf left panel} shows the map of \chisqmin\ derived from StarFISH over a wide range of distance moduli and extinctions. The continous line delimits the  $3\sigma$ (99.7\%) confidence level area of the best-fitting solution, while the dashed line delimits the $1\sigma$ (68\%) confidence level. The {\bf right panels} shows the best fitting solution in the form of the SFR in units of $\Msun\,{\rm yr}^{-1}$ as a function of the logarithm of age (blue histogram, top panel), for all age bins represented in the partial models, together with the averaged metallicity (black dots, middle panel). In the middle panel, the small green dots mark the central \mh\ and \logt\ values for which partial models are defined. In both panels, the vertical error bars mark the random errors in the SFR$(t)$ and $\mh(t)$ relations, while the grey-shaded areas delimit the systematic errors. 
}
\label{fig_sfr}
\end{figure*}
 
Table~\ref{tab_amr} displays the location of our partial models on the \feh\ vs.\ \logt\ plane. Age bins are defined as in \citet{Rubele_etal12}; they are, in general, at equally spaced intervals of $\Delta\logt=0.2$~dex, so as to reflect the fact that our age resolution is better at young ages. For the very youngest age bins, however, wider age intervals are used, so as to increase the statistics in places where the star counts are naturally very small. 

A total of 5 partial models are created for every age bin. They have metallicities covering an interval of $\Delta\feh=\pm 0.3$~dex with respect to the reference metallicity. The latter is derived from the \citet{p11} AMR, which results from observations of SMC star clusters. Metallicity bins are separated by 0.15~dex from each other. It is important to remark that the \citet{p11} AMR is used just to define the location of partial models, but it has virtually no effect in determining the final AMR which will be derived from our data (see Sect.~\ref{sec:amrdiscussion} later). 

To this set, we add one partial model representing the Milky Way foreground. This is generated by running the TRILEGAL Milky Way model \citep{Girardi_etal05}\footnote{\url{http://stev.oapd.inaf.it/trilegal}} for the central coordinates of each subregion, and for the total area it represents. 

We also note that partial models are computed for the entire range of distance moduli and extinctions explored in this work, namely $18.6\!<\!\dmo/{\rm mag}\!<\!19.3$ and $0\!<\!\av/{\rm mag}\!<\!0.8$.
At the high end of this distance/extinction ranges, the 10-Gyr main sequence turn-off might be located about 0.5~mag fainter than the $\ks<20.5$~mag limit being adopted in the derivation of the SFH. Anyway, this magnitude interval includes a good fraction of the subgiant branch for all ages, hence ensuring a good sensitivity to the old SFH. Moreover, we recall that the constraints on the oldest SMC populations do not depend on the main sequence and subgiant branches only, but also on their contribution to the RGB, RC, and horizontal branch, as demonstrated by \citet{Dolphin02}, and tested in the preparatory work for VMC by \citet{Kerber_etal09}.

Finally, it is worth mentioning that all models used in this work were produced using the Vegamag definition of photometric zeropoints, according to the procedure detailed in \citet{Girardi_etal02}. We seek for a good consistency between this definition and the zeropoints derived during the calibration of VMC data, as discussed in Appendix~\ref{sec:zeropoints}. For the moment, we just warn the reader that there might exist residual offsets between these two realisations of VISTA zeropoints, that might primarily appear as small offsets in the values of distances and extinctions derived in this work.

\subsection{Overview of the results for the SFR$(t)$ and $\mh(t)$}
\label{sec:sfhresults}

\begin{figure*}
\resizebox{0.95\hsize}{!}{\includegraphics{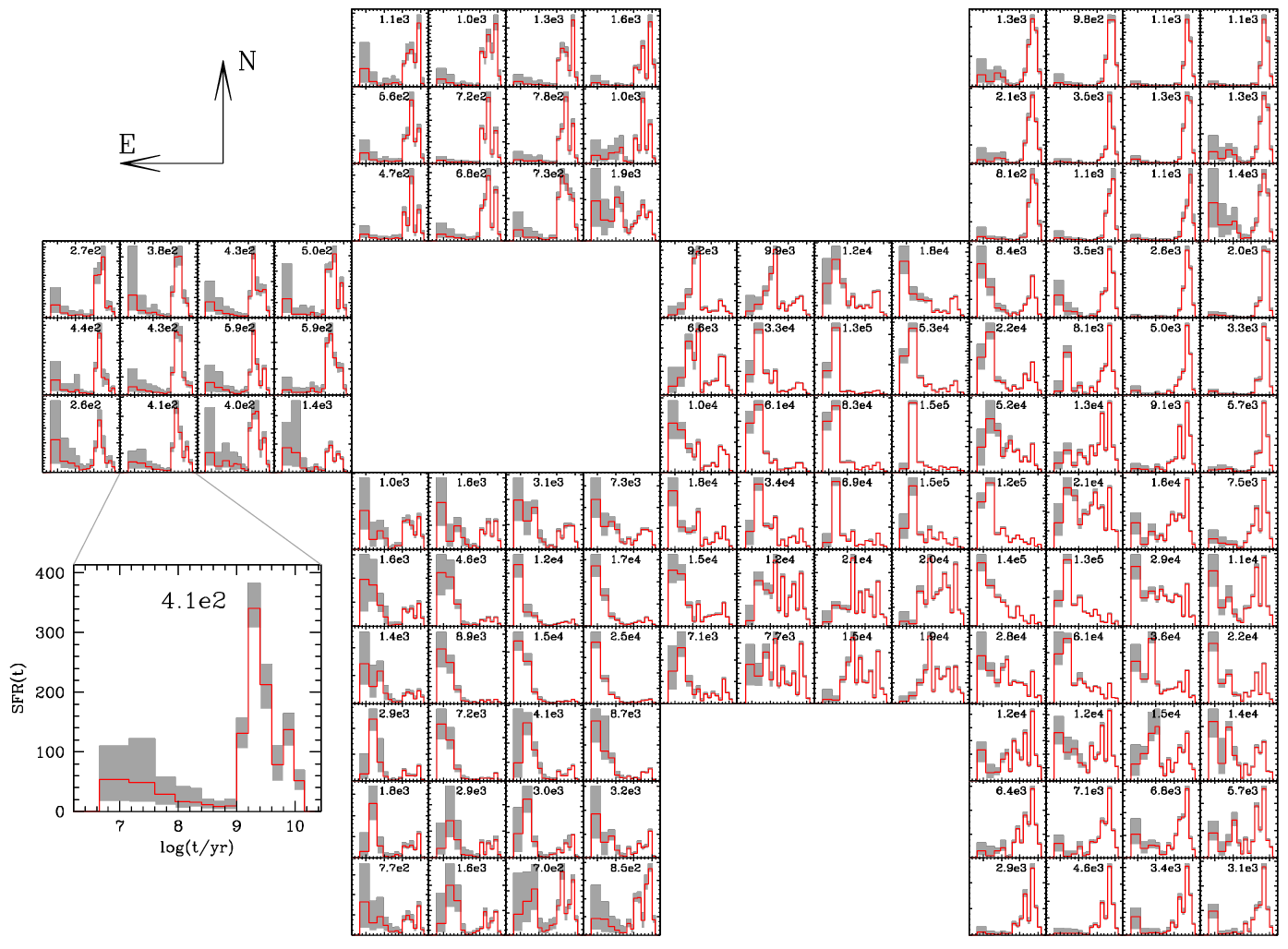}} 
\caption{An overview of the SFR$(t)$ for all subregions analysed in this work. The small panels present both the best-fit SFR$(t)$ (red line), and the $1\sigma$ confidence level (shaded grey area) for every subregion. They are displayed in tabular form, in the same way as they appear in the sky (see Fig.~\ref{fig_obstiles}). The axis limits are not presented in every small panel, because this would make the plot too crowded; anyway the axis limits can be easily derived with the help of the enlarged panel at the bottom-left, which serves as an example: All panels have the same limits in the ordinate, i.e. $6.2<\logt<10.5$. The limits for the SFR in the abscissa wildly vary between panels: they are set between 0 and a number which is presented inside each panel, with two significant digits; this number is in units of $10^{-7} \Msun\,{\rm yr}^{-1}$. }
\label{fig_sfr_east}
\end{figure*}

\begin{figure*}
\resizebox{0.95\hsize}{!}{\includegraphics{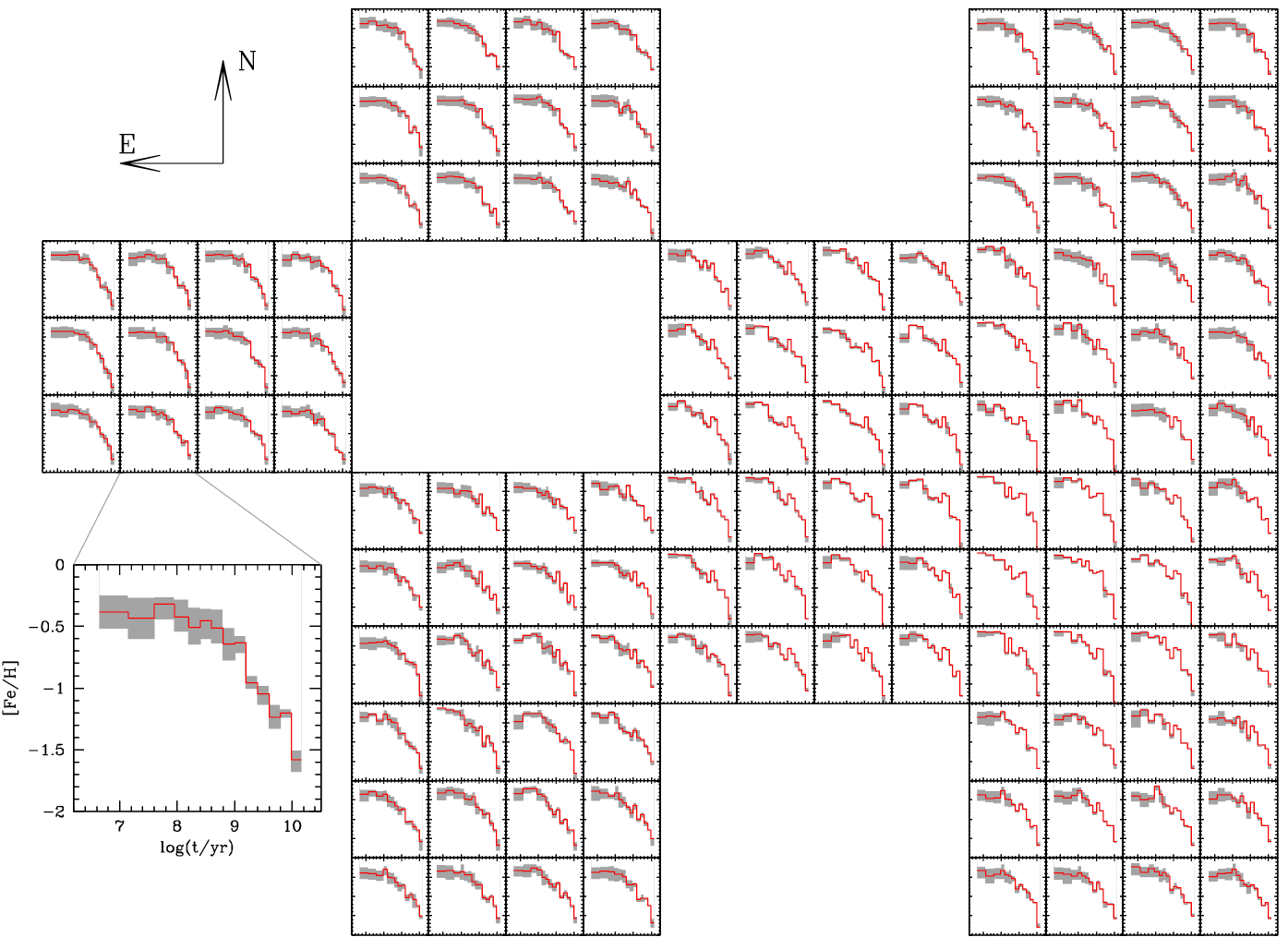}} 
\caption{The same as Fig.~\ref{fig_sfr_east}, but for the AMR, $\feh(t)$. This time, all panels present the same limits, as illustrated in the enlarged panel at the bottom-left.}
\label{fig_sfr_west}
\end{figure*}

\begin{table*}
\caption{Mean distances and extinctions for each subregion. A complete version of this table is provided in electronic form on the journal website.}
\label{tab_D_Av}
\centering
\begin{small}
\begin{tabular}{cc|cc|ccccccccc}
\hline
\hline
Tile & subre- &
$\alpha$ & $\delta$ & 
$\mu_0$ & $\sigma_{\mu_0,{\rm low}}$ & $\sigma_{\mu_0,{\rm up}}$ & 
$d$ & $\sigma_{d,{\rm low}}$ & $\sigma_{d,{\rm up}}$ & 
$A_V$ & $\sigma_{A_V,{\rm low}}$ & $\sigma_{A_V,{\rm up}}$  \\
 SMC     &   gion G      &
         (deg) & (deg) & 
         (mag) & (mag) & (mag) & 
         (kpc) & (kpc) & (kpc) &
         (mag) & (mag) & (mag)  \\
\hline
3\_3 & 1 & 13.18529695 & $-$74.59449494 & 18.92 & 0.10 & 0.08 & 60.81 & 2.68 & 2.81 & 0.426 & 0.051 & 0.099 \\
3\_3 & 2 & 11.94330634 & $-$74.61137246 & 18.90 & 0.07 & 0.08 & 60.26 & 2.01 & 2.08 & 0.454 & 0.079 & 0.046 \\
3\_3 & 3 & 10.59062191 & $-$74.60080543 & 18.97 & 0.10 & 0.08 & 62.23 & 2.78 & 2.91 & 0.382 & 0.057 & 0.043 \\
3\_3 & 4 & 9.23135534 & $-$74.58694737 & 19.02 & 0.07 & 0.08 & 63.68 & 2.09 & 2.16 & 0.369 & 0.044 & 0.031 \\
3\_3 & 5 & 13.27320506 & $-$74.20450581 & 18.93 & 0.08 & 0.07 & 61.09 & 2.11 & 2.18 & 0.409 & 0.035 & 0.040 \\
3\_3 & 6 & 11.95494707 & $-$74.20545383 & 18.98 & 0.05 & 0.07 & 62.52 & 1.54 & 1.58 & 0.416 & 0.041 & 0.084 \\
 
\hline
\end{tabular}
\end{small}
\end{table*}

The next steps leading to the final best-fitting SFHs are illustrated in Fig.~\ref{fig_sfr}, which refers to the analysis of subregion G8 of tile SMC 5\_4. Essentially, we run the StarFISH SFH-recovery software over a wide-enough grid of distance modulus and extinction values, \dmo\ and \av, as illustrated in the left panel. For each point in this grid, the code returns not only the best-fitting coefficients for the SFH, but also the \chisqmin. Its minimum value allows us to identify the best-fitting $(\dmo,\av)$ pair, which in this case falls at $\dmo=18.89$~mag and $\av=0.53$~mag. The coefficients of this best-fitting solution contain all the information necessary to express the SFH in the age--metallicity plane. For a more convenient representation of the results, for each age bin we derive the summed coefficients, producing the star formation rate function SFR$(t)$ in units of $\Msun\,{\rm yr}^{-1}$, and the average metallicity, hence deriving the age--metallicity relation, $\mh(t)$. They are illustrated in the right panel of Fig.~\ref{fig_sfr}.

We also evaluate the random and systematic errors in the SFR$(t)$ and $\mh(t)$ functions. One hundred simulations are produced using the TRILEGAL tool and the best-fitting solution, and then degraded using the ASTs and re-analysed by the SFH-recovery software. The scatter among these 100 solutions gives the SFH random errors. The distribution of the \chisqmin\ among these simulations allows us to identify the \chisqmin\ values that correspond to the $1\sigma$ and $3\sigma$ (68 and 99.7 \%, respectively) confidence levels. Systematic errors instead are evaluated by simply examining the total range of all the SFH solutions inside the $1\sigma$ area of the $(\dmo,\av)$ plane. The detailed procedure has been more extensively described in \citet{Kerber_etal09} and \citet{Rubele_etal12}.
  
Results for the 120 subregions analyzed in this paper are presented in Figs.~\ref{fig_sfr_east} and \ref{fig_sfr_west}. In these figures, we present the SFR$(t)$ and $\feh(t)$ resulting for all tiles and subregions, displaying them in tabular form, in the same way as they appear in the sky. The figures illustrate clearly the strong correlation between the SFH results for nearby subregions {\em and} tiles, especially for ages larger than $\logt\ga9$. The continuity in the SFH between neighbouring tiles is particularly encouraging, since -- contrary to subregions of the same tile -- they do not necessarily share the same photometric quality and completion in the \ks-band (see Table~\ref{tab_tiles}). 

Another feature evident from Fig.~\ref{fig_sfr_east} is that the SFR$(t)$ for the outermost SMC regions is overall quite similar, as can be appreciated comparing panels in the extreme corners of the surveyed area. This ``external'' SFR$(t)$ is composed of a main period of star formation between $\logt=9.2$ and $10.0$, plus a trace of star formation at young ages. In most of the external subregions, however, the young SFR is not significant compared to the error bars, so that we have only upper limits to it. Fields across the main SMC bar and on the Wing, instead, have a much more complex SFR$(t)$, with seemingly different bursts of star formation appearing along their histories, including very significant events of young star formation. Moreover, we can see that what appears as a single period of steady SFR for all ages $\logt>9.2$ in the extreme NW and SW of the survey area, in the Eastern most remote areas appears more like a double-peaked SFR$(t)$, with maxima at $\logt=9.5$ and 9.9. These two peaks are more evident in the NE tile SMC 6\_5. These features will be discussed further in Sect.~\ref{sec:agediscussion} below.
  
\section{Discussion}
\label{sec:close}
 
Although we obtain best-fit solutions and confidence intervals simultaneously for all parameters -- SFR$(t)$, $\mh(t)$, \dmo\ and \av\ -- it is convenient to discuss them separately in the following, postponing the general discussion of all trends to the final subsection.
 
\subsection{The SMC geometry}
\label{sec:geometry}
 
As already anticipated in the introduction (see also Sect.~\ref{sec:depth} below), several authours have derived significant depths along the SMC lines-of-sight. We, instead, have used the standard approach of assuming a null depth in the SFH analysis. This approximation might have two different implications. First, one may wonder whether the best-fitting distance and SFH we derive for each subregion are representative of the mean distance and SFH that would have been derived if a significant depth was adopted. This has been already verified by \citet{HZ04}, who simulated populations with depths up to 12~kpc (0.2~mag) and then analysed them with StarFISH without accounting for the spread in distance. They conclude that the SFH solutions were the same within the errors, and moreover that there were no indications of a significant depth in their results for the SMC. Since we are using the same software and method, we essentially endorse their conclusions and start assuming no depth {\em a priori}. The second implication of our null-depth approximation is that we will be constrained to analyse the SMC geometry more in terms of a ``thin surface projected on the sky'' rather than in terms of a complex 3D distribution. Changing this approach requires a dramatic change in the methods used in the SFH analysis, which will be pursued in subsequent works.

\begin{figure*}
\resizebox{0.48\hsize}{!}{\includegraphics{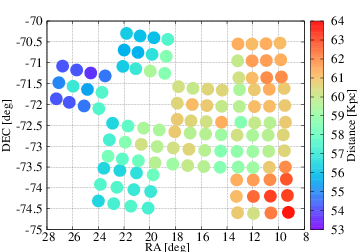}}\hfill
\resizebox{0.48\hsize}{!}{\includegraphics{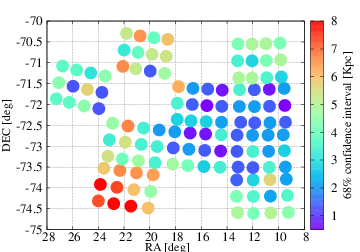}}
\caption{Maps showing the best-fit distances (left panel) and the widths of its 68\% confidence intervals (right panel) for each subregion. }
\label{fig_distmap} 
\end{figure*}

\begin{figure*}
\resizebox{0.48\hsize}{!}{\includegraphics{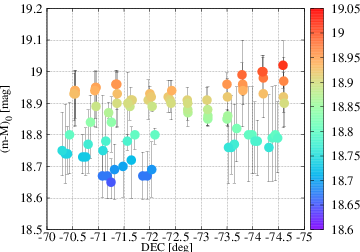}}\hfill
\resizebox{0.48\hsize}{!}{\includegraphics{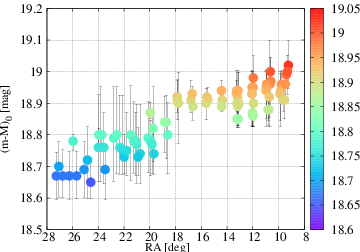}}
\caption{Distances and their 68\% confidence intervals as a function of coordinates $\alpha$ and $\delta$. It is evident that the distances increase systematically with $\alpha$, whereas there is no clear trend with $\delta$.}
\label{fig_distradec}
\end{figure*}

That said, the best-fitting and 68\% confidence intervals of the distances and extinctions are presented in Table~\ref{tab_D_Av} (a complete table is provided in electronic format on the journal website).

All distances are plotted as a function of equatorial coordinates in Figs.~\ref{fig_distmap} (left panel) and \ref{fig_distradec}, which indicate significant variations in its mean distance across the SMC and especially along the E--W direction. While central parts of the galaxy are found at distances between 58 and 60~kpc, the tiles SMC 3\_3 and 6\_3 are clearly farther out at $\sim\!62$~kpc, while the easternmost tile SMC 5\_6 -- the one closest in the sky to the LMC, and possibly associated with the Bridge -- is found at a distance of $\sim\!54$~kpc, which is more comparable to that of the LMC rather than the SMC. As can be appreciated in Fig.~\ref{fig_distradec}, there is a clear trend of increasing distances in the E--W direction. 

\begin{figure}
\resizebox{\hsize}{!}{\includegraphics{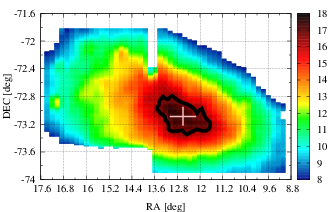}}
\caption{Stellar density map of the central SMC, where stars are selected as described in Sect.~\ref{sec:intro}. The black contour line is the density limit inside which the density variation is less than 10\% away from the peak value; this region is used to estimate the SMC stellar-density centre (pink cross), located at $\alpha=12.60^\circ$, $\delta=-73.09^\circ$ and $\dmo=18.856$~mag ($d=59.04$~kpc).}
\label{fig_findcentre}
\end{figure}

\begin{figure*}
\resizebox{0.49\hsize}{!}{\includegraphics{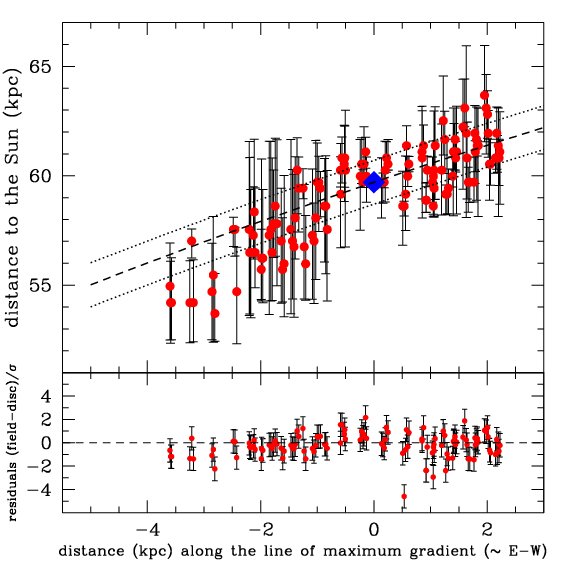}} \hfill
\resizebox{0.49\hsize}{!}{\includegraphics{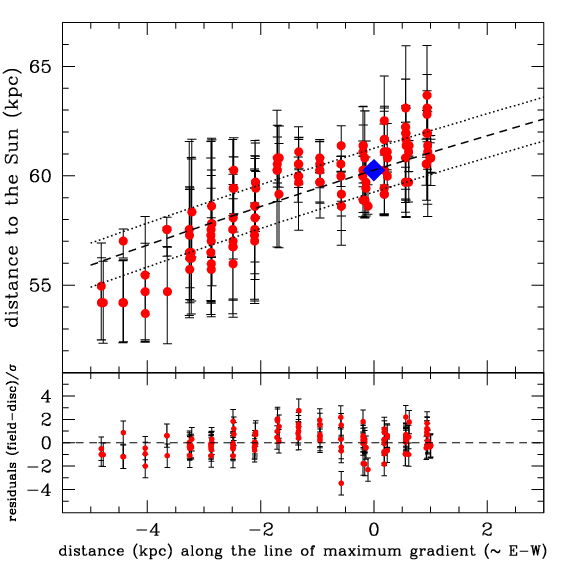}}
\caption{The two upper panels show the distance of the subregions as projected perpendicularly to the line of nodes, i.e. along the line of maximum distance gradient, which is close to the E--W direction. This line (dashed) is determined from fitting a plane to the distance data, for two different choices of the SMC centre (marked by a blue diamond): adopting either the kinematic (S04, left panels) or stellar density centre (this work, right panel). The dotted lines represent a projected disk thickness of 1~kpc (or 0.82~kpc linear distance) to both sides of the best-fitting plane. The bottom panels show the residuals with respect to best-fit plane, expressed in units of the $1\sigma$ uncertainty in the distance determination.}
\label{fig_fitplane}
\end{figure*}

We translate this E--W trend of the distances into a simple disk model as in the case of the LMC \citep[see][]{Rubele_etal12}. This was performed for two different choices of the SMC centre: the apparent kinematic centre from \citet[][S04]{stanimirovic04}, with $(\alpha_c\!=\!16.25^\circ, \delta_c\!=\!-72.42^\circ)$, and the stellar density centre indicated by the VMC data. The latter corresponds to $(\alpha_c\!=\!12.60^\circ, \delta_c\!=\!-73.09^\circ)$, and was determined as illustrated in Fig.~\ref{fig_findcentre}. We remark that it almost coincides with the centre of the K and M giants determined by \citet{god09}, with $(\alpha_c\!=\!12.75^\circ, \delta_c\!=\!-73.10^\circ)$. 
A weighted least-squares fit of a plane to the distance data then provides a position angle of the line of nodes of $\theta_0=179.5\pm1.3^\circ$ and an inclination angle of $i=39.7\pm5.5^\circ$ in the case of the S04 centre\footnote{$\theta_0$ and $i$ are defined as in \citet{vdM01}, i.e.\ $\theta_0$ is measured starting from the north in counter-clockwise direction, and the near side of the SMC is at $\theta_0-90^\circ$.}, and $\theta_0=179.3\pm2.1^\circ$ and $i=39.3\pm5.5^\circ$ in the case of our centre. 
The distances to the kinematic and stellar-density centres' intersection with the best-fit plane are  $\dmo=18.87\pm0.01$~mag ($59.48\pm0.28$~kpc) and $\dmo=18.91\pm0.02$~mag ($60.45\pm0.47$~kpc),   
respectively. These two cases are illustrated in Fig.~\ref{fig_fitplane}. 
Residual distances from these best-fit planes are reasonably high, amounting to up to $\sim\!2$~kpc in both directions. However, it is only in the easternmost regions that the approximation of a disk clearly fails, with residuals approaching $\sim\!-\!3$~kpc. Moreover, it can be noticed that at about 1.5~kpc to the east of the SMC stellar centre (which coincides with the kinematic centre) the best-fit plane is located systematically in front of the data points by about $1$~kpc. These deviations are generally larger than the $1\sigma$ errors in the distance determinations, as can be appreciated in the bottom panels of Fig.~\ref{fig_fitplane}. The general picture we get is that the SMC is very distorted, and if its structure is to be approximated by a disk, it is a significantly warped one.
Another interesting aspect is that the position angle of the line of nodes of the best-fit plane appears not to be related with the major axis along which most of the SMC young stars and gas are distributed (i.e.\ the so-called SMC bar), which has a position angle of $\sim40^\circ$, and which also coincides with the kinematic position angle \citep[cf.][]{stanimirovic04}. Similar results have been obtained recently by \citet{tatton14}, from the analysis of the RC magnitude in VMC data.

It should be stressed that the choice of fitting a disk is not determined by a belief that the SMC has (or is) a disk; it is mainly driven by two facts: (1) it is the simplest surface that can be fit to a table of distances as a function of coordinates, as the one which we have derived; and (2) because finding the mean disk parameters allows us to easily compare our distance results with a series of other similar derivations of the SMC ``best-fit plane'' available in the literature. The two main parameters we find, namely the inclination and position angle, could also be interpreted in terms of other simple 3D configurations like, for instance, oblate or prolate spheroids.

Regarding the comparison with other results in the literature, it is generally found that both the old and intermediate age populations (ages $>\!2$~Gyr) are distributed in a pressure supported, spheroidal/ellipsoidal distribution \citep[][and references therein]{sub12}. The young stars, like classical Cepheids, are found to be in a disk. In the present work we do not use any specific distance tracer, but the use of near-infrared data implies that our distances are generally more weighted by the intermediate-age populations than by young stars. So, it is initially a surprise that we find indications of a distorted disk, rather than the more smooth features that would be expected from a spheroidal/ellipsoidal distribution. Moreover, it also surprising that the structure we \citep[and also][]{tatton14} obtain is very different from the one derived from the classical Cepheids, which indicate a much more inclined disk with $i\sim65-70^\circ$ \citep[][]{caldwell86}. Cepheids also possess a very different position angle of line of nodes, $PA\sim145 - 150^\circ$ \citep[see also][]{laney86, Haschke_etal12}.

\subsection{Possible depth structures}
\label{sec:depth}

The total distance intervals corresponding to the 68\% confidence level of the best-fit distances are illustrated in the right panel in Fig.~\ref{fig_distmap}. One can see clearly that such widths are larger towards the SE side of the SMC, and especially at the SE border of tile SMC 3\_5, which has a distance confidence interval about $8$~kpc wide. Most of the other fields have best-fit distances constrained to within $5$~kpc -- with a few fields having distances constrained to be inside $\lesssim1$~kpc. One can also notice that there is a mild trend of the total distance interval increasing gradually towards the SE of the SMC.

The interpretation of this increased distance confidence level intervals is not straightforward, but two possible causes for them are: (1) indications that these areas present large distance spreads along the line of sight, or (2) that the smaller stellar density towards these outer regions are causing sitematically poorer fits of the SFHs and distances. Indeed, \citealt{Kerber_etal09} have shown that the SFH errors are primarily a function of stellar density and completeness. However, it is striking that the tiles SMC 5\_6 and 6\_5, towards the NE, present similar stellar densities (see Fig.~\ref{fig_obstiles}) and smaller completion than tile SMC 3\_5 (see Table~\ref{tab_tiles}), but no indication of an increased distance confidence level interval. This lead us to suspect that the SE areas of the SMC might indeed present significant depths along the line of sight. This suspicion cannot be confirmed at the moment, since our method does not assume a distance dispersion. A subsequent work will be dedicated to this point.
 
In this context, however, it is worth recalling that many other authours have found indicated distance variations and/or a sustantial depth across different lines-of-sight of the SMC, starting with the early work by \citet{gardiner91}. The literature of the topic is highly varied and based on very heterogeneous data. The latest works on the subject, however, tend to use the RC position and widths as provided by major surveys such as the Magellanic Cloud Photometric Survey (MCPS) and the Optical Gravitational Lensing Experiment (OGLE), as the primary probe of the SMC geometry. Depths derived in this way often reach values between 10 and 23 kpc \citep{gardiner91, sub09n, nidever13, tatton14}.

Interestingly, using the position and width of the RC, \citet{nidever13} find that the eastern side of the SMC has a bimodal distribution of distances, with a component located at just 55~kpc, far closer than the main SMC body which they locate at 67~kpc. Our best fit-distances (left panel in Fig.~\ref{fig_distmap}) essentially confirm the presence of this close component of the SMC, which is clearly seen in tile SMC 5\_6 (Fig.~\ref{fig_distmap}, left panel) for typical distances of $56$~kpc. However, our distance maps do not produce evidence for the farther component (expected at $>60$~kpc) at this position. This could simply be caused by the nearby component being denser than the farther one, so being the only one being singled out by our method. Indeed, examination of the CMDs for the SMC tile 5\_6 (Fig.~\ref{YKcmd}) reveals the presence of an extended RC, with the faintest half being less populated than the brightest; this latter probably corresponds to \citet{nidever13}'s distant component. 

Finally, before closing the discussion related to the distances, it is worth recalling that the distances determined with our method are average values obtained assuming that in any subregion stars of all ages are at the same mean distances. The most representative ages sampled by this method vary a lot across the SMC, as we will see in Sect.~\ref{sec:tomo}, but it is fair to state that they are most sensitive to the intermediate-age populations that (if present) generate prominent RCs and RGBs in our near-infrared CMDs. However, there are clear indications from other authors that populations of different ages have different geometries. For instance, based on the large sample of variables from OGLE~III \citet{Haschke_etal12} find that the young classical Cepheids are distributed with a position angle of the major axis $\theta=66\pm15^\circ$ and inclination $i=74\pm9^\circ$, while the very old RR~Lyrae variables are better described by a $\theta=83\pm21^\circ$ and $i=7\pm15^\circ$. Both distributions are clearly different from the mean geometry we find. This is not surprising  for the young populations, which are concentrated on the SMC bar. The different results of the very old population, instead, seem to indicate that this old population constitutes an extended halo \citep[see also][]{nidever_etal11} with little resemblance to the warped disk galaxy we derive.
 
Needless to say, future analysis of the SFH based on VMC data will include a more detailed description of the distance distributions along the different lines-of-sights, and as a function of stellar ages. In this context, the present solutions constitute a valuable first-guess for this more complex analysis. Moreover, we are also studying the structure of the SMC as traced by classical Cepheids and RR~Lyrae using the whole sample from OGLE III plus EROS-2 variables which have a counterpart in VMC, and new period--luminosity relations derived from the multi-epoch VMC data \citep[see e.g.][]{Ripepi_etal12,ripepi14,Moretti14}. This will enable us to make a much more robust comparison between the geometry derived from the variables and from the SFH-recovery work.

\subsection{Large-scale extinction maps}
\label{sec:extinctionmap}

\begin{figure}
\resizebox{\hsize}{!}{\includegraphics{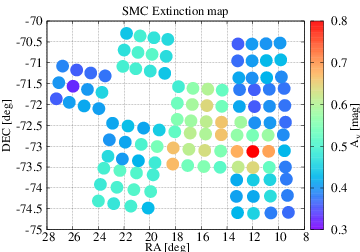}}
\caption{Map of best-fit extinctions \av.}
\label{fig_avmap}
\end{figure}

The extinction map derived from our best-fitting values of \av\ is presented in Fig.~\ref{fig_avmap}.  As can be appreciated, most of the external SMC regions are found at extinctions typically smaller than $\av=0.5$~mag, with minimum values being of $\av\simeq0.35$~mag. Only across the main bar, and in parts of the Wing regions in tile SMC 3\_5, are the \av\ values larger than this, reaching $\av>0.65$~mag for three subregions. The average extinction for the entire analysed area is of $\av=0.47\pm0.09$~mag. We remark that the maximum value of $\av=0.65$~mag corresponds to extinctions of just 0.25, 0.18 and 0.07~mag in the filters $Y$, $J$ and \ks, respectively, and to colour excesses of 0.18~mag in \yk\ and 0.11~mag in \jk\ (cf.\ item~\ref{item_alav} in Sect.~\ref{sec:method}). 
 
The comparison with extinction values from the literature is not easy, given the very different methods and probes applied by different groups. The widely-used extinction maps from \citet{Schlegel_etal98}, based on estimating dust column densities from dust emission, are not reliable for the central parts of the SMC; for its immediate surroundings, however, they indicate mean values close to $\av\!\sim\!0.12$~mag, which probably represent the typical values for the foreground MW extinction across the entire SMC. Similar dust maps were derived for the central SMC regions by \citet{israel10}, using COBE/DIRBE and WMAP data; they provide a mean extinction of about $\av=0.45$~mag {\em internal to the SMC}. That means that their total {\em average} extinction values reach $\av\simeq0.57$~mag, which compares very well with our results for the SMC subregions with the highest mean extinctions. Overall, however, \citet{israel10} extinctions appear larger than our values. This could be explained by the fact that these extinction maps are derived from emission-based estimates of the total dust column density, hence comprising the entire line-of-sight to the SMC galaxy, while our extinction values are a star-by-star weighted mean that reflect the mean values somewhere in the middle of the SMC stars.

On the other hand, our average \av\ values agree quite well with those derived by \citet{zar02} from MCPS data, which span the range of $0.15-0.65$~mag, when considering both cool and hot stars. In their case, the extinction is probed using similar objects (the stars), but using a different method, namely a star-by-star fitting of the spectral energy distribution.

\citet{Haschke_etal11} instead find an average extinction of just $\av=0.1\pm0.15$~mag (or $\evi=0.04\pm0.06$~mag) adopting a theoretical mean unreddened colour for the RC stars, and $\av=0.18\pm0.15$~mag (or $\evi=0.07\pm0.06$~mag) from RR~Lyrae stars from the OGLE~III survey. A similar distribution of the extinction in the central part of the SMC was found in \citet{sub12}, also using the mean colour of the red clump. In both cases, the derived extinction values appear to be significantly smaller than ours. We should however keep in mind that these are mainly measuremements of the {\em internal} SMC extinction, that is, the RC method is mostly sensitive to the variations of extinction with respect to a fixed reference value. Moreover, there is a significant difference in assumptions with respect to the present work: in practice, both \citet{Haschke_etal11} and \citet{sub12} assume that the RC has a constant intrinsic colour across the SMC. We, instead, assume that all CMD features (including the RC) vary as indicated by stellar models and following the age and metallicity distributions of the subregions being sampled. Tracing back the origin of differences in the final \av\ values, in these circumstances, is not straightforward.

However, if we consider the two least-extincted northeastern tiles (namely tiles SMC 5\_6 and 6\_5), there indeed appears to be a modest zero-point offset of $\sim\!0.2$~mag in our \av\ values, compared to the canonical 0.12~mag derived from \citet{Schlegel_etal98} for the SMC outskirts. We note that such an offset, if real, could be easily explained by offsets of the order of $\sim\!0.05$~mag in the VMC photometric zeropoints, and/or in the stellar models employed here.
Moreover, \citet{peek10} show that differences in dust temperature may lead to variations in the \citet{Schlegel_etal98} $E_{B\!-\!V}$ values of a few hundredths of a magnitude (a tenth of a magnitude in \av), so that it is not even clear whether the problem is in our results or on the dust emission maps. Although this surely deserves further investigation, we recall that such errors would not affect other aspects of this work, since they would, prevalently, just cause a small systematic offset in our distance moduli (that is, a $\Delta\dmo\simeq0.05$~mag). Indeed, all comparisons in the distances and reddenings we are making here are chiefly meant to explore how these quantities vary across the SMC in a {\em differential} way. Future works by the VMC team will better address the problem of determining the absolute values of distances and extinctions from these data.


\begin{figure}
\resizebox{0.48\hsize}{!}{\includegraphics{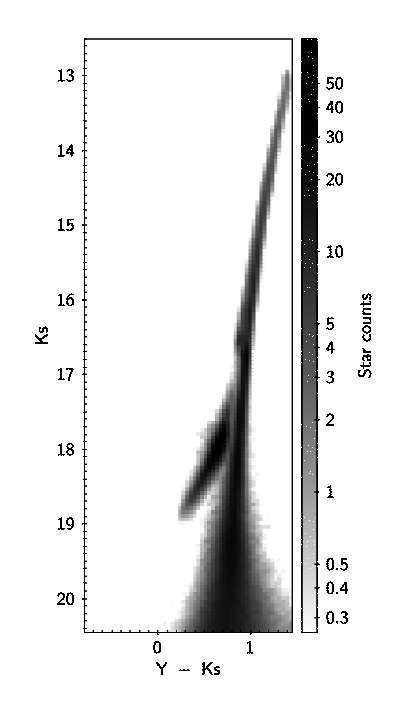}}\hfill
\resizebox{0.48\hsize}{!}{\includegraphics{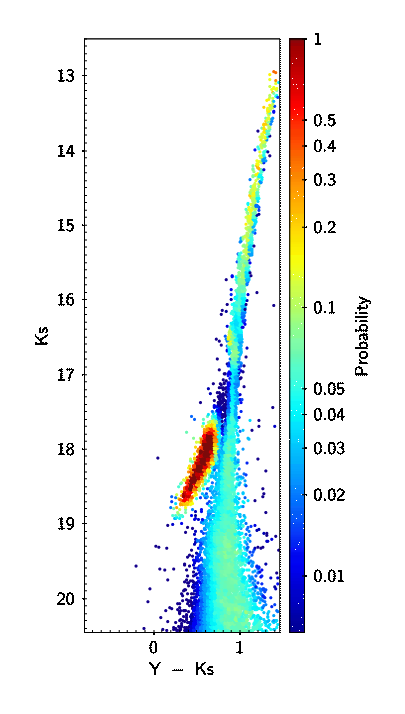}}
\\
\resizebox{0.48\hsize}{!}{\includegraphics{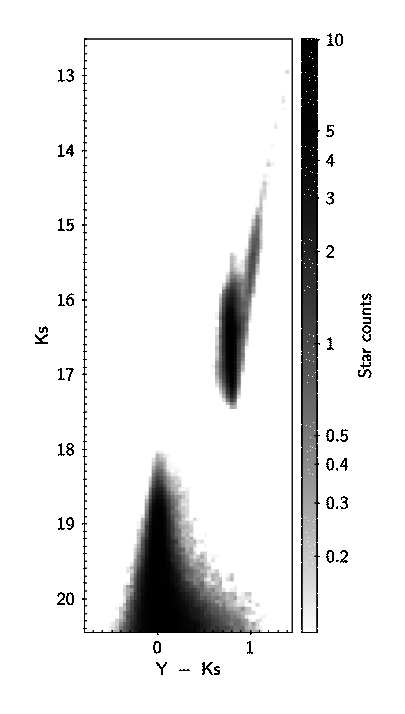}}\hfill
\resizebox{0.48\hsize}{!}{\includegraphics{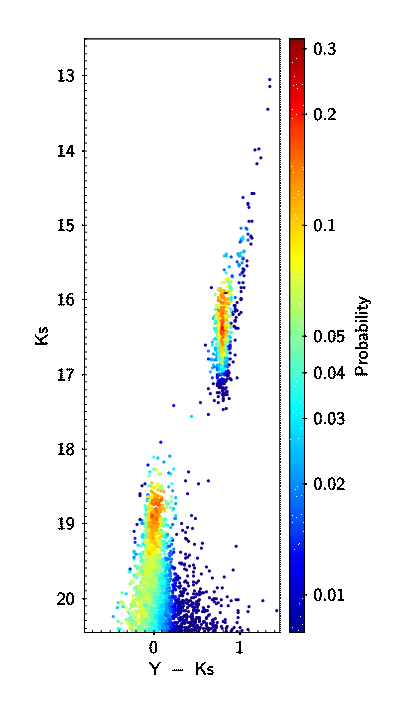}}
\caption{An intermediate step in the Stellar Populations CMD Reconstruction. The top-left panel shows the Hess diagram of the best-fit partial model of a subregion in tile SMC 3\_3 with $\logt=10.075$ and $\feh=-1.45$~dex. The top-right panel instead shows a simulated stellar population (dots) of exactly the same age and metallicity, but with the individual stars being coloured according to the probability of them belonging to that specific stellar population -- as opposed to the probability of belonging to all the other partial models that have been fit to this subregion. The bottom panels shows the same but for a young stellar population, with $\logt=8.7$ and $\feh=-0.50$~dex, in the same subregion.}
\label{fig_spr}
\end{figure}

\begin{figure*}
\resizebox{0.3114\hsize}{!}{\includegraphics{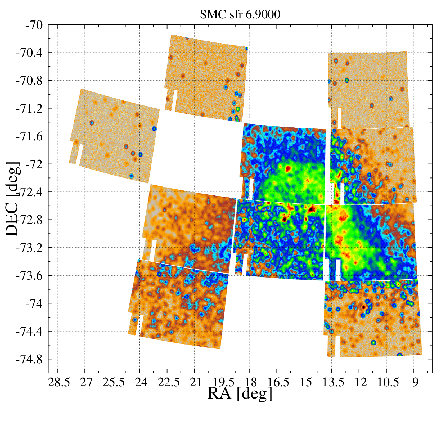}}
\resizebox{0.3114\hsize}{!}{\includegraphics{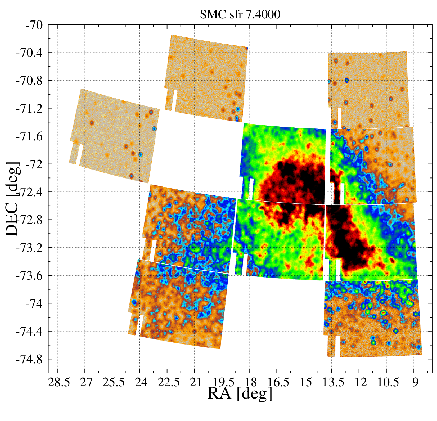}}
\resizebox{0.3600\hsize}{!}{\includegraphics{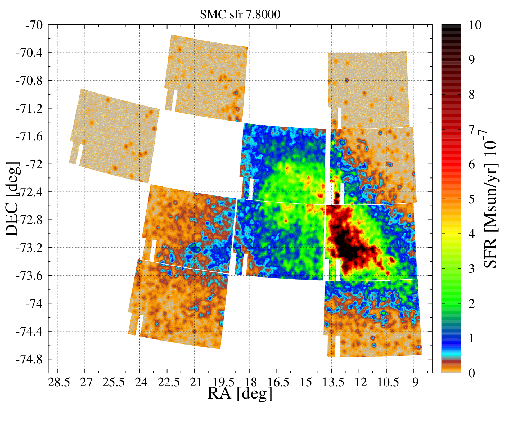}}\\
\resizebox{0.3114\hsize}{!}{\includegraphics{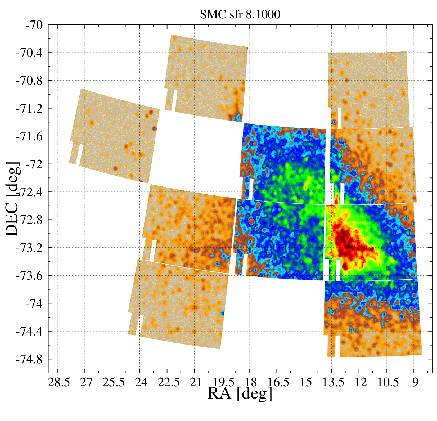}}
\resizebox{0.3114\hsize}{!}{\includegraphics{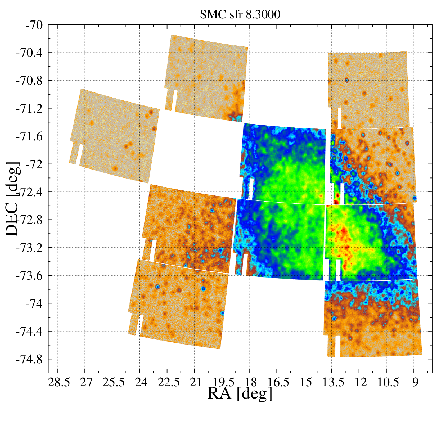}}
\resizebox{0.3600\hsize}{!}{\includegraphics{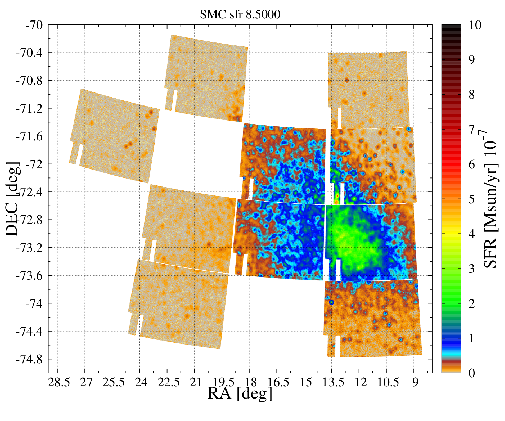}}\\
\resizebox{0.3114\hsize}{!}{\includegraphics{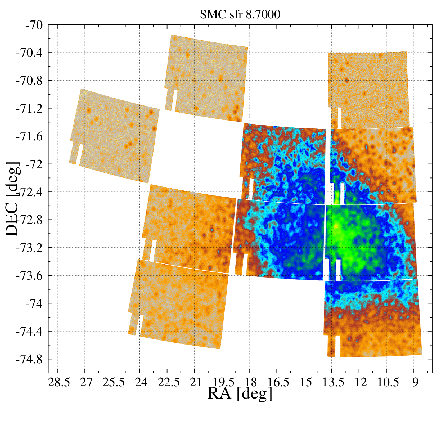}}
\resizebox{0.3114\hsize}{!}{\includegraphics{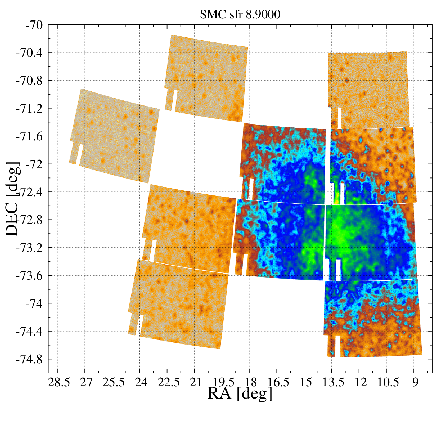}}
\resizebox{0.3600\hsize}{!}{\includegraphics{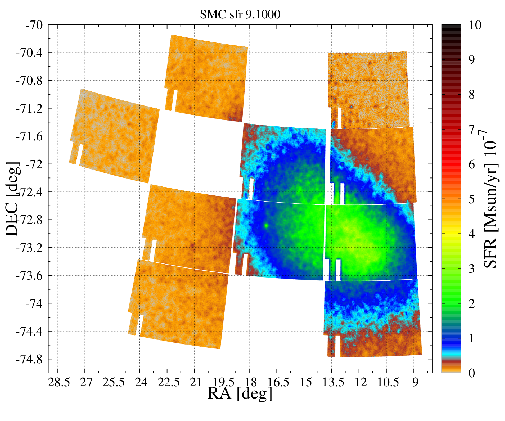}}\\
\resizebox{0.3114\hsize}{!}{\includegraphics{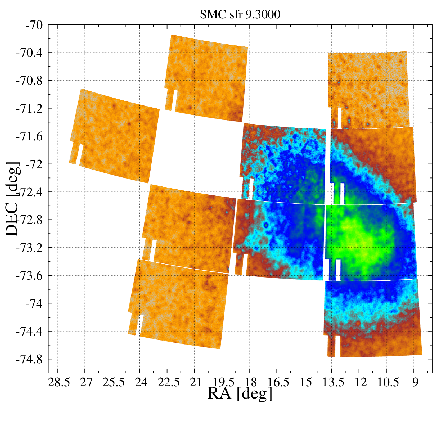}}
\resizebox{0.3114\hsize}{!}{\includegraphics{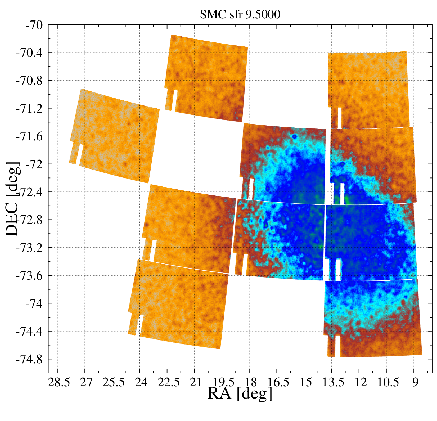}}
\resizebox{0.3600\hsize}{!}{\includegraphics{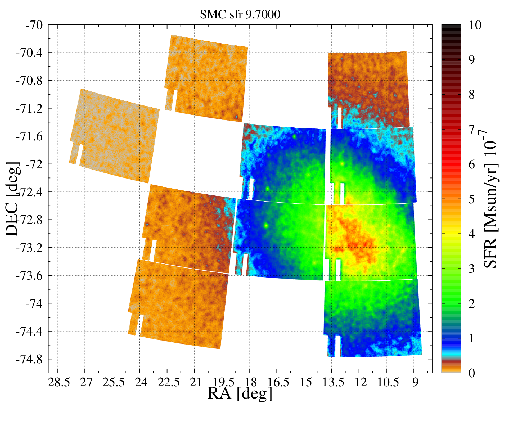}}
\caption{Detailed SFR$(t)$ maps. Each panel shows the SFR intensity for a given age bin, derived as detailed in the main text. The mean \logt\ of each age bin is indicated above the top axis; they coincide with the values listed in Table~\ref{tab_amr}. \label{fig_tomosfr} }
\end{figure*}
\begin{figure*}
\resizebox{0.3114\hsize}{!}{\includegraphics{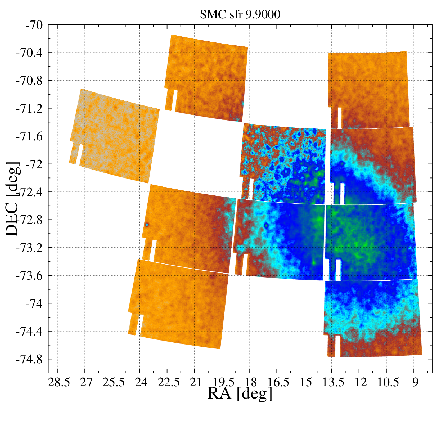}}
\resizebox{0.3600\hsize}{!}{\includegraphics{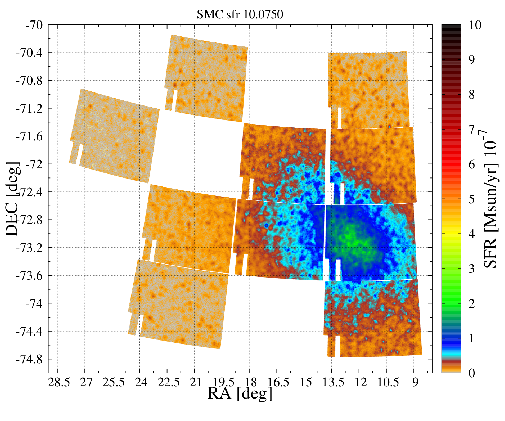}} \\
{\bf Figure~\ref{fig_tomosfr}} {\it continued}.
\end{figure*}

\subsection{The look back time star formation rate and mass assembly tomography}
\label{sec:tomo}

Figures \ref{fig_sfr_east} and \ref{fig_sfr_west} encode a lot of information about the SFH of the SMC, including their spatial distribution and derivations of the uncertainties. However, they do not provide a good visualisation of the relative spatial distribution of the different populations; moreover, those are low-resolution maps whose pixel elements are equal to the subregions. In order to improve upon this point, we have transformed our SFH results, together with the original photometry data, into high-resolution maps of the SFH and history of the SMC stellar mass assembly.

Our stellar population maps are built as follows:
\begin{enumerate}
\item For each one of the $120$ subregions analysed, we take the coefficients of the best-fitting solution and build the total Hess diagram of each age bin $i$ and metallicity $j$.
\item For each star in the photometric catalogue, we identify its cell in the Hess diagram. The total number of stars in that cell is $n_{\rm obs}$. 
\item We assign to the star the relative probabilities of belonging to each one of the $14\times5$ age--metallicity bins, by simply computing the ratio 
\[
p_{ij} = n_{ij,{\rm bfm}} \times \frac{ n_{\rm obs} }{ \sum_{ij} n_{ij,{\rm bfm}}}
\]
where $n_{ij,{\rm bfm}}$ is the number found in the best-fit model.  
\item We sum the $p_{ij}$ of the five metallicity bins holding the same age $i$, hence deriving $p_i$.
\item We randomly assign the star to an age bin, following the probability set by the distribution of $p_i$. 
\item Finally, we build density maps containing all stars which have been observed. There is a map for each age bin.
\end{enumerate}

An intermediate step of this process is illustrated in Fig.~\ref{fig_spr}, which shows how a population of a given age and metallicity (that is, stars belonging to a single partial model) would appear in terms of $p_{ij}$. The top half of the plot shows how stars belonging to an old age appear, when translated into a probability of being truly old -- or, more specifically, of belonging to the partial model with $(i,j)=(14,1)$: they appear very prominently ($p_{14,1}>0.3$) if they are at the faintest part of the red clump and horizontal branch, and moderately prominent ($0.05<p_{14,1}<0.2$) over the RGB and reddest part of the turn-off, while stars at the bluest part of the turn-off turn out to be given a much lower probability of being truly that old.  For a young population, instead, largest probabilities of being truly young are in general assigned to the stars at the brightest part of the He-burning sequence, and at the bluest part of the turn-off, as illustrated in the bottom panels of Fig.~\ref{fig_spr}, for a population with $(i,j)=(7,3)$. In our method, the same probability assignment is made to all stars observed by VMC. It is evident that the resulting population maps will heavily weight the stars whose ages, due to their particular position in the CMDs, have a high probability of not belonging to other age bins.

The resulting spatial maps are presented in Fig.~\ref{fig_tomosfr}, for all 14 age bins we have defined. As the previous Fig.~\ref{fig_sfr_east}, they also reflect the intensity of the best-fitting SFR$(t)$, but with a significant advantage: across each subregion, this SFR$(t)$ is spatially distributed in proportion to the number of stars that likely have that age, and the resolution of the SFR maps is significantly improved.  
Despite the presence of a few artifacts -- caused e.g.\ by the detector 16 gaps, or by the imperfect superposition between tiles -- a few features are remarkable in the maps of Fig.~\ref{fig_tomosfr}:
\begin{itemize}
\item populations of all ages older than $\logt\!>\!9.7$ present a nearly-elliptical distribution of their stars;
\item the elliptical shapes persist for ages between $9.1\!<\!\logt\!<\!9.7$, although at these ages the distribution seems to have a sharper decline in density at the NE end of the ellipse;
\item at younger ages, however, even more significant distortions appear, especially at the SW side of the SMC, which starts to clearly delineate the Wing for all ages $\logt\le 8.3$;
\item finally, at very young ages ($\logt<8.3$), a very patchy distribution of star formation appears, more concentrated in the SMC bar and Wing. A prominent and wide episode of star formation is evident at the NE extremity of the bar at the $\logt=7.4$ age bin. This region is well known from previous optical surveys and contains the prominent young cluster NGC~346.
\end{itemize}
 
The presence of young stars in the Wing starting at $\logt<8.3$ ($t<0.2$~Gyr), is consistent with previous simulations which are based on the LMC-SMC-Galaxy interaction \citep{yoshizawa03,yozin14}. These simulations clearly show that the formation of young stars in the Wing is due to the last LMC-SMC interaction about 0.2~Gyr ago, when the outer gas disk was tidally compressed during the very strong tidal LMC--SMC interaction (with the distance of the two being $\sim\!10$~kpc). Our observations appear to confirm the LMC--SMC-interaction origin of the Wing formation.

Our maps can be compared with the large-scale population maps derived by \citet{HZ04} from MCPS data. Such maps were obtained with a very different resolution, and probably reflect many of the differences in the data quality and implementation of the SFH-recovery method. Although the comparison is not easy, some clear similarities are evident at first sight, especially in regard to the distribution of the young star formation. 

However, an aspect worth mentioning is that we do not find evidence of the large ring of star formation found by \citet{HZ04} at ages of 2.5~Gyr, suggesting that this feature might have been an artifact derived from their significantly shallower photometry than ours, added to the uncertainties intrinsic to the optical work -- such as for example the larger sensitivity to differential extinction of MCPS.

\begin{figure}
\resizebox{\hsize}{!}{\includegraphics{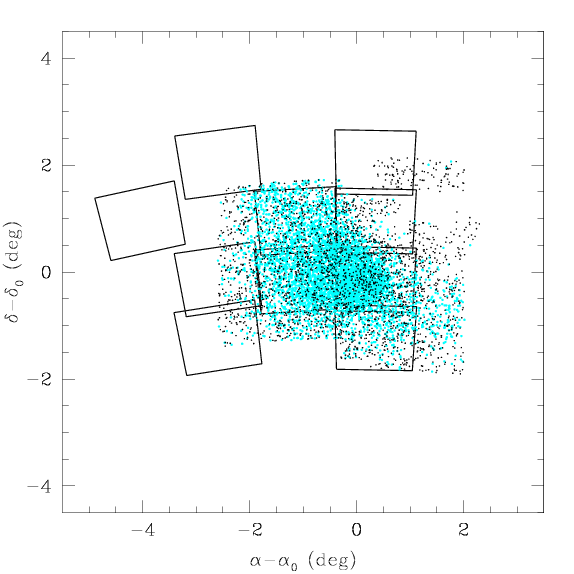}}
\caption{Distribution of classical Cepheids (cyan points) and RR Lyrae stars (black points) in the SMC. There are 4630 Cepheids from OGLE-III plus 165 from EROS-2, and 2475 RR Lyrae stars from OGLE~III plus 133 from EROS-2. The black boxes are the VMC tiles analyzed in this work.}
\label{fig_variables}
\end{figure} 
 
These maps compare well with those inferred from the distribution of classical Cepheids (tracing the young, $t\!<\!100$~Myr, population) and RR~Lyrae stars (tracing the old, $t\!\simeq\!10$~Gyr, population) in the SMC. This is shown in Fig.~\ref{fig_variables}, where cyan points are classical Cepheids and black points are RR Lyrae stars. Although OGLE~III and EROS-2 cover only part of the area analyzed in this paper (specifically, entirely tiles: SMC 5\_4, 4\_4, 5\_3, 4\_3 and 3\_3,  half of the tile 4\_5, and $\sim\!1/4$ of tiles 3\_5 and 6\_3)
it is quite clear that the density of classical Cepheids peaks in tiles SMC 5\_4, 4\_4 and 4\_3, which are dominated by the young population, according to our SFR$(t)$ maps. On the other hand, the RR~Lyrae stars are distributed rather smoothly all over the SMC area -- although in more central SMC tiles many of them probably do not show up because ``buried'' amid the overwhelming young population.

\begin{figure*}
\resizebox{0.3114\hsize}{!}{\includegraphics{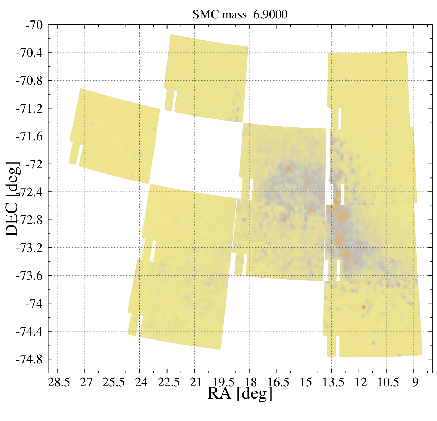}}
\resizebox{0.3114\hsize}{!}{\includegraphics{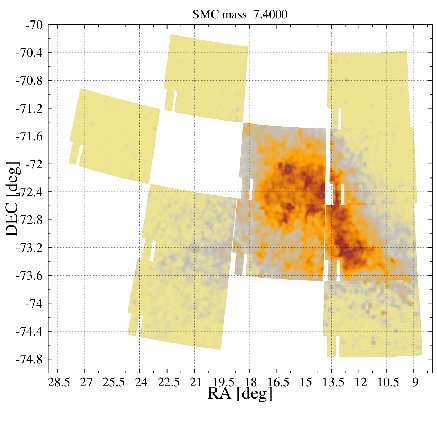}}
\resizebox{0.3600\hsize}{!}{\includegraphics{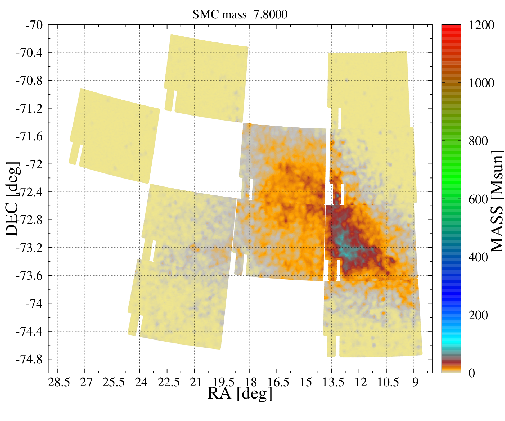}}\\
\resizebox{0.3114\hsize}{!}{\includegraphics{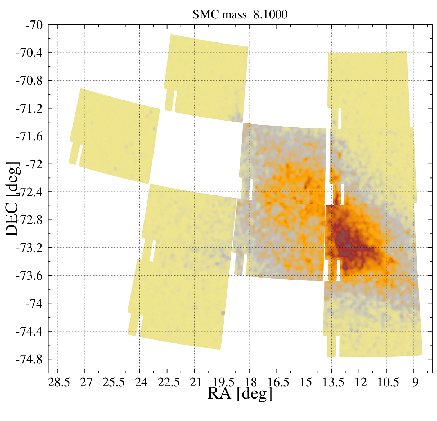}}
\resizebox{0.3114\hsize}{!}{\includegraphics{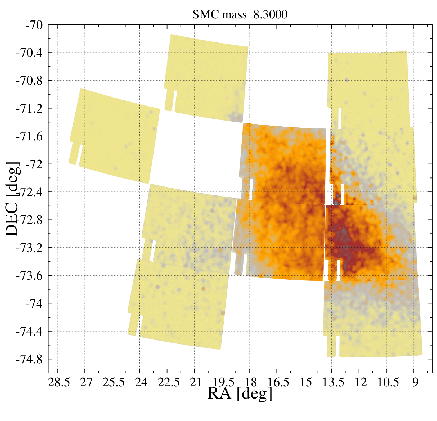}}
\resizebox{0.3600\hsize}{!}{\includegraphics{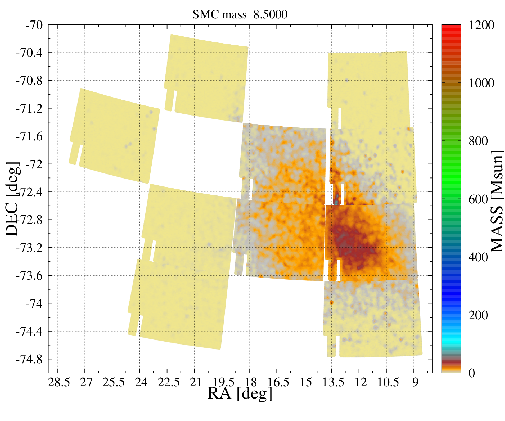}}\\
\resizebox{0.3114\hsize}{!}{\includegraphics{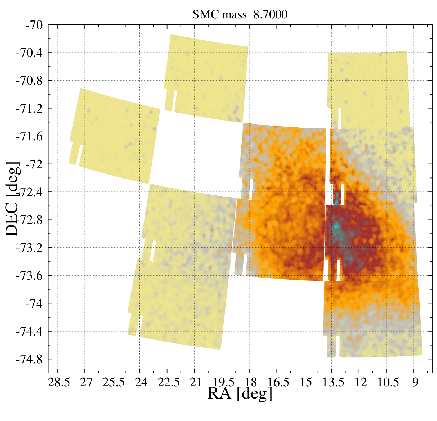}}
\resizebox{0.3114\hsize}{!}{\includegraphics{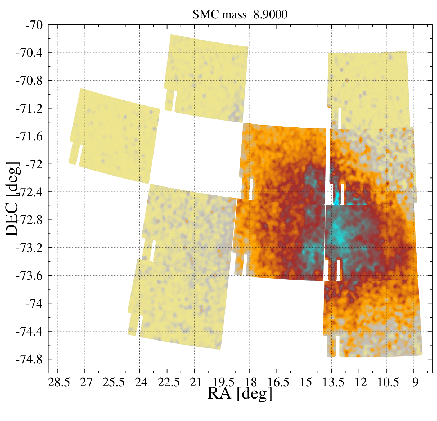}}
\resizebox{0.3600\hsize}{!}{\includegraphics{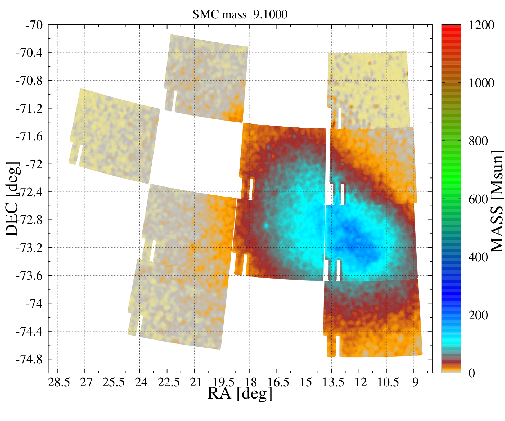}}\\
\resizebox{0.3114\hsize}{!}{\includegraphics{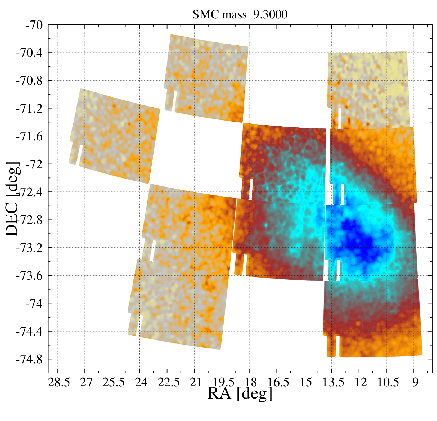}}
\resizebox{0.3114\hsize}{!}{\includegraphics{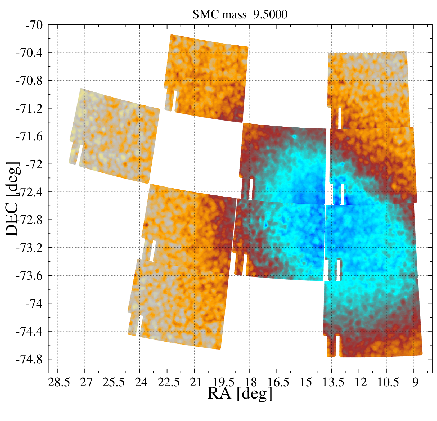}}
\resizebox{0.3600\hsize}{!}{\includegraphics{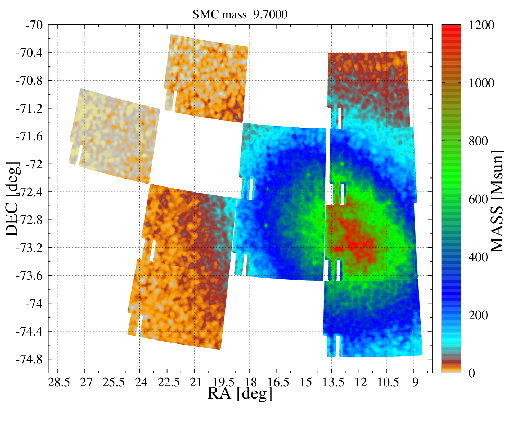}}
\caption{The same as in Fig.~\ref{fig_tomosfr}, but now on a colour scale that reveals the total stellar mass formed in each age bin. \label{fig_tomomass} }
\end{figure*}
\begin{figure*}
\resizebox{0.3114\hsize}{!}{\includegraphics{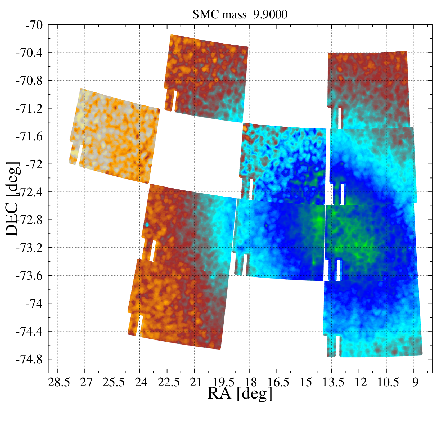}}
\resizebox{0.3600\hsize}{!}{\includegraphics{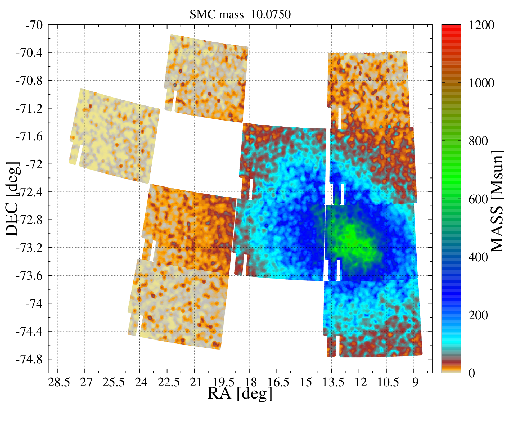}} \\
{\bf Figure~\ref{fig_tomomass}} {\it continued}.
\end{figure*}

Since the SFR$(t)$ maps are built for bins almost-equally spaced in $\logt$, they do not reflect the total mass of stars built in each epoch of the SMC history. To derive the latter, we redo the maps re-scaling them to the total stellar mass formed in each age bin. The results are presented in the maps of the stellar mass assembly of Fig.~\ref{fig_tomomass}. As can be seen, these maps reveal that the young star formation constitutes only a minor fraction of the total stellar mass formed in the SMC. It is remarkable however that the $\logt=9.7$ age bin ($4.0$ to $6.3$ Gyr, or $\sim\!5$~Gyr) turns out to contain a significant fraction (34~\%) of the total stellar mass formed in the SMC. 

This latter result is in agreement with \citet{rezaei14}, who, analysing the $K$-band luminosity function of long period variables found/classified by \citet{soszynski11} and \citet{ita04a}, also find the main epoch of stellar mass assembly in the SMC to have occurred at around 6~Gyr ago, at a rate of $\sim\!0.28~\Msun\,{yr}^{-1}$.
 
\subsection{The global SFR$(t)$}
\label{sec:agediscussion}

\begin{figure}
\resizebox{\hsize}{!}{\includegraphics{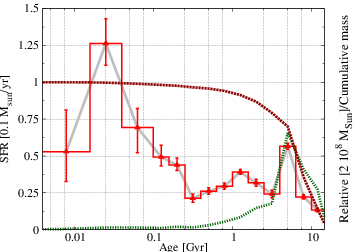}}
\caption{The SMC global SFR$(t)$ (solid red line), together with its error bars. Also shown are the relative stellar mass formation as a function of age (green dotted line), and the cumulative stellar mass normalised to the total formed mass (of $3.85\times10^8$~\Msun; red dotted line). }
\label{fig_GSFR}
\end{figure}

With the detailed SFR$(t)$ and population maps to hand, we can perform a rough comparison with the SFR$(t)$ derived by other authors. Before that, we build the SMC global SFR$(t)$ as illustrated in Fig.~\ref{fig_GSFR}. Considering the large coverage of the central parts of the SMC provided by the present VMC data, it is likely that this global SFR$(t)$ describes most of the stellar mass formed by this galaxy.

Figure \ref{fig_GSFR} shows a few periods of more intense star formation in the SMC. Essentially, it tells us that the SFR$(t)$ was relatively modest at very old ages, with mean values below $0.015\Msun\,{\rm yr}^{-1}$. SFR intensified significantly at an age of $\sim5$~Gyr, reaching a peak of $0.05\Msun\,{\rm yr}^{-1}$; during this period, a large fraction of the stellar mass of the SMC was formed. At intermediate ages, the SFR$(t)$ declined to values as low as $0.025\Msun\,{\rm yr}^{-1}$, but with another peak of $0.04\Msun\,{\rm yr}^{-1}$ at $\sim1.5$~Gyr. Finally, more age resolution is provided at young ages, for which it is apparent that the very recent star formation has increased by factors between 2 and 5 with respect to the values dominant at intermediate-ages. The young SFR$(t)$ peak is found in the second age bin, spanning from 14 to $40$~Myr. Figure \ref{fig_GSFR} also shows the cumulative stellar mass formed since the oldest ages. It turns out that $3.85\times10^8$~\Msun\ of stars were formed in the area presently sampled by VMC; this estimate obviously depends on the adopted initial mass function, which comes from \citet{Chabrier03}.
  
Comparing our results with the literature is not easy, considering the  significant number of studies in the past, based on very different datasets and methods. The era of quantitative SFH derivation starts in practice with a small region with available HST/WFPC2 data analysed by \citet{Dolphin_etal01}. It was followed by the impressive work by \citet{HZ04}, based on the MCPS, which marked the field for years. Although the MCPS resolution was not comparable to the HST one, it had the great advantage of completely covering the SMC main body. The main findings from \citet{HZ04} were the presence of several periods of enhanced star formation, with one in particular (at $2.5$~Gyr) causing a large-scale annular structure in the SMC. As already mentioned, we do not find evidence of such a structure.

In more recent years, different sets of deep photometric data were used for the same goal of deriving the field SFH \citep[e.g.][]{chiosi07,netal09,cetal13,wetal13}; although they have reached a general consensus regarding some basic SFH features of the SMC, they have also been dissimilar to some extent. 
For ages $>12$~Gyr, the SFH seems to be consistent with a relatively low star formation, as judged from the CMD analyses of deep HST photometry of seven spatially diverse fields by \citet{wetal13}, as well as with ground-based $BR$ photometry of twelve distinct fields spread across the galaxy by \citet{netal09}.
According to \citet{netal09}, such a low old SFR is similar at all radii and azimuths. No region inside $4.5$~kpc from the SMC centre has been found to be dominated by the old population.
These findings by \citet{netal09} and \citet{wetal13} broadly agree with our conclusions, that the SFR has been modest at all ages larger than $5$ Gyr. 

Relatively old and intermediate-age peaks have also been found in the SMC SFH at ages of $\sim\!4.5$ and $9$ Gyr, and suggested to be the result of tidal interaction with the LMC \citep{wetal13}. In the western fields studied by \citet{netal09}, the older enhancement splits into two peaks, located at $\sim\!8$ and $\sim\!$12~Gyr. However, HST $VI$ CMDs of six other fields located at projected distances of $0.5$ to $2$ degrees from the SMC centre resulted to exhibit a slow star formation pace since the galaxy formation until $\sim\!5$~Gyr ago, when the star formation activity started to increase sharply \citep{cetal13}. Such an increase peaks at $\sim1.5-2.5$ Gyr. Additionally, the level of the intermediate-age SFR enhancement systematically increases towards the centre. Whether gas from the outer regions has been centrally funneled still remains an issue, due to uncertainties in the geometry of the SMC \citep{wetal13}. As far as we are aware, the conundrum about the existence of several peaks at ages older than $2-3$~Gyr depended on the unavailability of datasets for more areas across the SMC. In regard to these issues, our data clearly confirms (1) the peaks of star formation at ages of $1.5$ and $5$~Gyr, and (2) that the peak of SFR at $1.5$~Gyr appears, effectively, more concentrated than the $5$~Gyr one, as can be appreciated in the population maps of Fig.~\ref{fig_tomosfr}.
A similar result was obtained by \citet{rezaei14}, who finds that the SMC experienced two main episodes of star formation at $\sim\!6$~Gyr ($\logt=9.8$) and $\sim\!0.7$~Gyr ($\logt=8.8$), being the latter more concentrated in the central parts.

Noteworthily, the $1.5$-Gyr old peak of star formation finds correspondence in a peak of cluster formation occurring at about the same time in the SMC, LMC, and the MW \citep{piatti10, piatti11a, piatti11b}. This coincidence might be indicating a close encounter with the Milky Way at these ages, although such an event is strongly disfavoured by the most recent proper motion data for the LMC \citep{besla07, kal13}.

As far as ages younger than $500$~Myr are concerned, the SFH is mostly driven to the SMC innermost regions, where the major star formation events took place. This age range is featured by several peaks of star formation which have preferentially occurred towards the eastern side of the galaxy \citep{netal09}. \citet{is11} also found evidence of a shift for the centroid of population younger than $500$ Myr and up to $40$ Myr in the direction of the LMC using the $VI$ OGLE-III and the MCPS. Our results broadly confirm these suggestions, but provide substantial improvements in the spatial resolution and area of the derived young SFH.
 
In summary, some features of our SFH find correspondence with the results in the literature, others do not:
\begin{itemize}
\item The initial period of low SFR is supported by the HST-based studies by \citet{wetal13} and \citet{cignoni12, cignoni13}, and by the ground-based study by \citet{netal09}.
\item In the western fields studied by \citet{netal09}, the older enhancement is found to split into two peaks, located at $\sim$8 and $\sim12$~Gyr. Although our data does not extend as far to the west as theirs, we do not find indications of a similar split, since our age bins imply a coarser resolution of the oldest SFR$(t)$ with respect to \citet{netal09}. 
\item The period of intense SFR at $5$~Gyr finds strong support in the HST-based work by \citet{wetal13}, who date this event at $4.5$~Gyr, and also in \citet{netal09} and \citet{cignoni12}.
\item \citet{HZ04} find bursts of star formation at $2-3$~Gyr, $400$ Myr, and $60$ Myr. The first and last of these periods seem to have correspondence in our data, although at slightly different ages.
\item \citet{cetal13} find an increased SFR peaking at $\sim1.5-2.5$ Gyr, which is consistent with suggestions of a close encounter with the Milky Way at recent time. We essentially find the same, at $1.5$~Gyr.
\item From a different kind of data, \citet{rezaei14} finds main episodes of star formation at $\sim\!6$~Gyr and $\sim\!0.7$~Gyr, at rates in reasonably good agreement with our data, and with a concentration of the younger star formation towards the center, similar to the trends we find.
\end{itemize}

As a final note, we point out that all the previous works on the SMC SFH seem complementary, in the sense that they analyse the SFH at different spatial scales and with very different sensitivities regarding the old SFH. From the point of view of the present work, they all suffer from the same problem: a higher sensitivity to the presence of differential extinction. None of these surveys can rival VMC in terms of spatial coverage, except for the MCPS; and yet, MCPS is the survey that more dramatically illustrated how the SFHs derived from optical data are sensible to differential extinction: \citet{HZ04} even had to resort to different amounts of extinction affecting red and blue stars, in order to improve their CMD fits. With its sensitivity to dust reduced by a factor of at least three, as compared with a survey based on $VI$ data, this is one aspect where VMC has probably made significant progress.

\subsection{The metallicity evolution} 
\label{sec:amrdiscussion}
\label{sec:amrreview}

In the context of the SFH-recovery work, the AMR can be treated either as an integral part of the derivation method, or as a by-product. The difference is in whether the AMR derived from field observations is used to force the SFH solution to occupy a certain line in the age--metallicity space, or whether the best-fit SFH solution is used to indicate which AMR actually holds. We have used the latter approach, allowing the AMR to vary within a wide area in the age--metallicity space, providing enough freedom for slightly different AMRs to be derived between different regions. Needless to say, such an approach would not have been possible were VMC not to cover an extremely large area of the SMC.

\begin{figure}
\resizebox{\hsize}{!}{\includegraphics{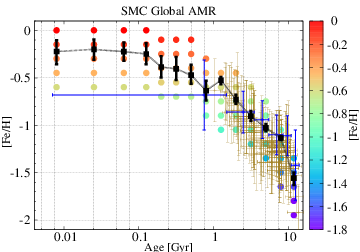}}
\caption{The SMC global AMR (solid black line with error bars). The colour points show the distribution of the stellar partial models in the AMR diagram. The olive points with error bars show the distribution of clusters analysed in \citet{p12}. The violet points are the mean values obtained by \citet{car08} from the combination of 13 SMC fields; in this case the vertical error bars indicate the metallicity dispersion inside the age bins delimited by the horizontal error bars.}
\label{fig_AMR}
\end{figure}

The several space-resolved AMRs have already been introduced in Fig.~\ref{fig_sfr_west}. In addition, we have produced the global AMR illustrated in Fig.~\ref{fig_AMR}, which is simply the weighted mean \feh\ computed over all of the SMC, for each age bin. The general features are clear: the \feh\ has generally increased in time, with the exception of a period of slight decrease in metallicity taking place for ages slightly younger than $1$~Gyr.

Several independent determinations of the AMR exist in the literature, most of them derived from the study of ages and metallicities in star clusters.
\citet{p11} used the largest known SMC cluster sample with ages $\ge1$~Gyr and metallicities put onto a homogeneous scale, and found two periods of enhanced cluster formation at $t\sim2$ and $5-6$ Gyr, taking place throughout the entire body of the galaxy. He also found the absense of a metallicity gradient, and a spread in metallicity for clusters older than $\sim7$~Gyr. 
A compromise between covering a relatively large area and reaching the oldest main sequence turn-offs was accomplished by \citet{p12}. He found that the field stars do not possess gradients in age and metallicity, and that stellar populations formed since $\sim2$~Gyr ago are more metal-rich than $\feh\sim-0.8$~dex and are confined to the innermost region (semi-major axis $\le$ 1$^\circ$).  He also compared the field star AMR to that of the star cluster population with ages and metallicities in the same field scales, and found that clusters and star fields share similar chemical evolution histories. This latter conclusion seems to be fully supported by the comparison with our results for the field, shown in Fig.~\ref{fig_AMR}. For comparison, the figure also presents the mean AMR obtained by \citet{car08} from caltium triplet observations in 13 different SMC fields. Note that there is a good agreement between the different AMRs for all ages larger than 2~Gyr. For the youngest age bin, the comparison is meaningless because the RGB-dominated sample by \citet{car08} does not provide age information for stars younger than $\sim1$~Gyr.

One of the most remarkable points in the global AMR of Fig.~\ref{fig_AMR} is the peak in metallicity at $1.5$~Gyr, followed by a decrease at $0.8$~Gyr, and then finally followed by a steady increase that leads to the present-day metallicities. This indicates an event of dilution in the metallicity having occurred $\sim1.2$~Gyr ago.

\begin{figure*}
\resizebox{0.32\hsize}{!}{\includegraphics{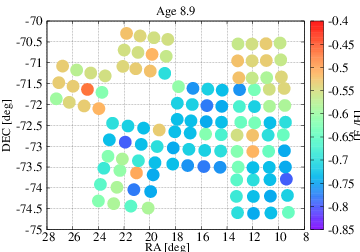}}
\resizebox{0.32\hsize}{!}{\includegraphics{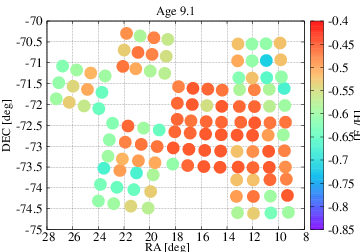}}
\resizebox{0.32\hsize}{!}{\includegraphics{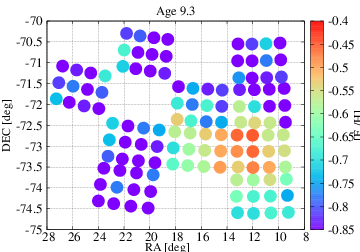}}
\caption{Spatial distribution of the mean metallicity derived for each age bin between $\logt=8.9$ and $9.3$ ($0.8$ to $2$~Gyr).}
\label{metmap}
\end{figure*}

In order to better understand this feature, in Fig.~\ref{metmap} we plot the spatial distribution of the metallicity in each sub region analysed, in the relevant age interval.
It is remarkable that the peak in metallicity manifests itself first in the age bin $\logt=9.3$ ($2$~Gyr), in the form of a higher metallicity in the inner SMC regions. For the age bin $\logt=9.1$ ($1.3$~Gyr), the entire SMC is observed to present a higher metallicity. Finally, for the age bin $\logt=8.9$ ($0.8$~Gyr), the metallicity has already decreased again, but for the Northern outskirts of the SMC -- which, coincidentally, are the only regions to not have formed stars recently. Although the dynamics of the SMC stellar populations is quite uncertain over such timescales, the effect appears as if a major phase of accretion of metal poor gas has taken place in the SMC history, interrupting the normal increase in the system's metallicity, at an age of 1~Gyr, so that the new star formation took place starting with a metal-poorer gas. 
Interestingly, the period of global decrease in the SMC metallicity follows the $1.5$-Gyr peak in the global SFR$(t)$. This concomitance might be indicating a major event in the SMC history, like for instance the merging with a gas-rich companion galaxy, or simply a closer passage to the LMC or the Milky Way.

Indeed, this latter interpretation seems consistent with the simulations by \citet{yozin14}, which predict that the LMC--SMC--Galaxy interaction could have triggered enhanced star formation in the SMC about 1.5~Gyr ago. This is also the epoch of the Magellanic Stream formation. They find that the 1.5-Gyr ago interaction can strongly distort the entire SMC body so that a large amount of gas in the outer part (thus metal-poor) can be funneled to the central bar region, where enhanced star formation can occur\footnote{Such simulations do not predict enhanced star formation at $\sim\!5$~Gyr, because this is beyond the scope of the currently existing LMC-SMC interaction models. A possible burst (via merger) around 7.5~Gyr ago in the SMC is discussed by \citet{tb09}. The 5~Gyr burst could constitute a useful constraint on such models, being described e.g.\ as the epoch of the first arrival close to the MW, or the epoch when the SMC started its strong interaction with the LMC.}.  When the outer metal-poor gas is transferred to the central region to be consumed by star formation,  the new stars should have lower \feh\ initially. However, owing to the chemical evolution after the burst, \feh\ should increase again. A subsequent study by \citet[][see their figure 14]{yozin14} does not show clearly the dip in the AMR, perhaps because they averaged the AMR for the entire region. 
In the present work, however, the same galaxy-wide average still causes the metallicity dip to be clearly seen in the data.
 
On the other hand, as a cautionary note, we recall that the $>\!1.5$~Gyr interval also corresponds to the ages at which the RGB quickly develops in stellar populations, so we cannot exclude that our method reacts to this sharp evolutionary transition by introducing some artificial feature in the AMR at those ages. However, in support of the ``galaxy-interaction'' hypothesis, is that most of the massive intermediate-age clusters in the LMC and SMC turn out to be observed exactly at ages close to $1.5$~Gyr \citep[see e.g.][]{Girardi_etal13}. Moreover, the most massive LMC and SMC clusters of ages between $1$ and $2$ Gyr are known to present multiple turn-offs and dual red clumps \citep[][]{Goudfrooij_etal09, girardi09}, which might be related to the same phenomenon: i.e., a galaxy interaction that has disturbed the ISM enough to drive the accumulation of gas in the cluster interiors, and the subsequent star formation there. Although this might sound as rather speculative, this picture could explain a wide range of observational facts, without resorting to extreme assumptions.

\section{Summary and closing comments}
\label{sec_summary}

In this work, we have provided SFH, distance and extinction maps across an area of 14~deg$^2$ of the SMC. The maps are computed with an initial resolution of 20\arcmin, which is the size of the 120 nearly-square subregions used in the SFH-recovery. Although maps of similar or better resolution have been presented before, our work is distinguished from them by basically three crucial aspects: (1) the use of uniform and high-quality near-infrared photometry, reaching stars near the oldest main sequence turnoff across all of the observed area and with a reduced sensitivity to differential extinction; (2) the simultaneous derivation of all the quantities from the best global fit of the CMDs; (3) the extremely wide area that has been analysed, in a homogeneous way. 

These results already allow us to revise previous results regarding the SMC morphology. We find that it is systematically tilted in the E--W direction, with a mean inclination angle of $39^\circ$. However, there are enough significant displacements with respected to an inclined disk, to allow us to infer that the SMC is significantly distorted and similar to a warped disk.
In addition, a large distance spread along the line of sight is clearly detected in the SE end of the SMC, which corresponds to the end of its Wing. 
After deriving the SFH for all these subregions, we assign to every observed star a probability of belonging to a given age--metallicity interval. This allows us to make high-resolution maps of the several stellar populations observed across the SMC. We can, for instance, illustrate the flocculant nature of the very young star formation, and identify the nearly smooth features drawn by the older stellar generations. 
We identify the centre of the intermediate-age to old stellar populations to be located at coordinates $(\alpha_c=12.60^\circ, \delta_c=-73.09^\circ)$, which is $3.7^\circ$ displaced with respect to the kinematical SMC center determined by \citet{stanimirovic04}. We document the formation of the SMC Wing at ages younger than $\sim\!0.2$~Gyr, and the dramatic features drawn by the young star formation in the SMC Bar, which peaked at ages of about $40$~Myr. Even more importantly in terms of the SMC history, we clearly detect periods of enhanced star formation at ages of $1.5$ and $5$ Gyr ago. The former is probably related to an interesting feature we find in the AMR, that points to the ingestion of metal-poor gas taking place at ages slightly larger than $1$ Gyr. The latter, instead, constitutes a major period of stellar mass formation in the SMC. Indeed, we confirm that the SFR was moderately low at even older ages in the SMC.

These findings are still to be further explored, and fully interpreted in terms of the possible scenarios for the evolution of the SMC. Although they clearly confirm many previous claims in the literature, they also provide significant more detail on several aspects, such as the wide-scale distributions of the populations with different stellar ages.   

The SFH data are available at \url{http://stev.oapd.inaf.it/VMC_SFHdata/}. 
The analysis of additional VMC data, that over the next four years will complete the SMC and LMC mosaic, is in progress.

\section*{Acknowledgments}

We thank the Cambridge Astronomy Survey Unit (CASU) and the Wide Field Astronomy Unit (WFAU) in Edinburgh for providing calibrated data products under the support of the Science and Technology Facility Council (STFC) in the UK.
RdG acknowledges partial research support from the National Natural Science Foundation of China (NSFC) through grant 11373010.  This work was partially supported by the Argentinian institutions CONICET and Agencia Nacional de Promoci\'on Cient\'{\i}fica y Tecnol\'ogica (ANPCyT). PM acknowledges the support from the  ERC Consolidator Grant funding scheme ({\em project STARKEY}, G.A.\ n.\ 615604).

\appendix
 
\section{Homogenized deep tiles and PSF photometry}
\label{sec:vmcdata}
  
\begin{figure*}
\resizebox{0.33\hsize}{!}{\includegraphics{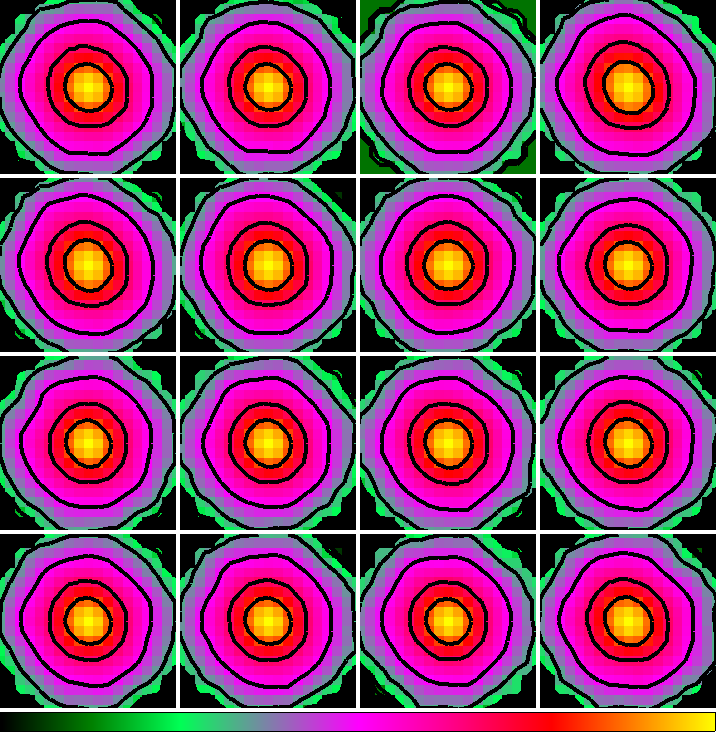}} \hfill
\resizebox{0.33\hsize}{!}{\includegraphics{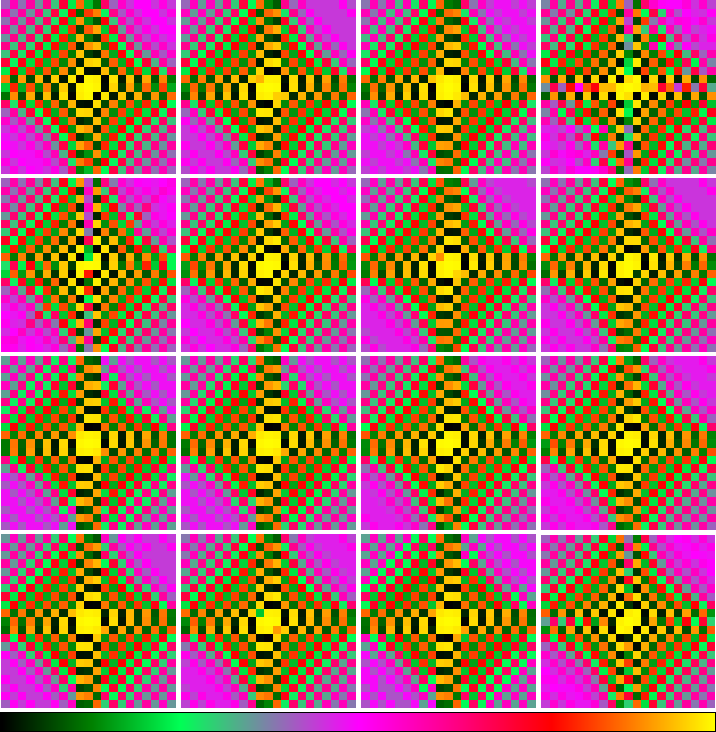}} \hfill
\resizebox{0.20\hsize}{!}{\includegraphics{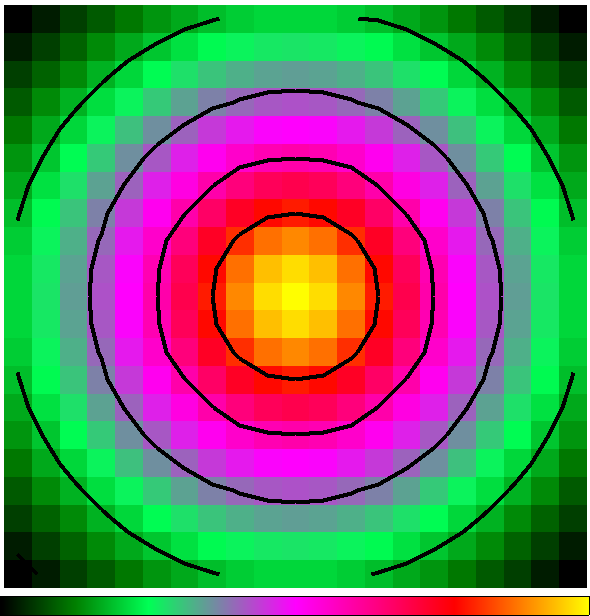}}
\caption{{\bf Left Panel:} An example of the derived PSF model on each detector on a pawprint image at a given epoch. {\bf Central panel:} The deconvolved kernels for the same detectors. {\bf Right panel:} The RPSF model used to produce the kernels.}
\label{psf_homog}
\end{figure*}

To generate a deep tile we combine images from different epochs (at least three in $Y$ and $J$ and twelve in the \ks\ band), each one formed by the combination of six pawprint images. These images are stacked single exposures reaching a total exposure time of $400$ seconds in $Y$ and $J$, and $375$ seconds in \ks, for each pawprint\footnote{The average observing time across the deep tile is about twice these values.}. Variations of the seeing occurring over these timescales could affect the PSF shape on the final deep tile, as a function of position.  To avoid this, we apply a homogenization method to correct the PSF on each detector, in each pawprint image, in each epoch, to a constant and homogeneous PSF model, before generating a deep tile. We proceed as follow:
\begin{enumerate}
\item First we derive the effective PSF model (EPSF) on pawprint images of a given epoch. To do that we assume a constant PSF across each detector of the VIRCAM camera, and use the IRAF/DAOPHOT tasks to find the best-fit PSF model. Usually, it corresponds to a Moffat function profile. Examples of such EPSFs are shown in the left panel of Fig.~\ref{psf_homog}.
\item Afterwards, we generate a symmetric Moffat function reference PSF model (RPSF) with a half-flux radius (HFR) equal to the largest HFR of the EPSF model of all detectors, in all pawprints images, at a given epoch.
\item Subsequently, using a Fourier deconvolution method we derive the kernels that, when convolved with the pawprint images, convert them to images with a homogeneous RPSF model. An example of kernels and its RPSF model, applied on a pawprint image are illustrated in the central and right panels of Fig.~\ref{psf_homog}.
\item Finally we combines all homogenized pawprint images with the SWARP tool \citep{Bertin02} in the same way as described in \citet{Rubele_etal12}, thus generating a deep tile image with a homogeneous PSF.
\end{enumerate}

With this method, we obtain epoch images (in both pawprint and tile formats) with a homogeneous PSF, of course degraded to the worst 
pawprint at a given
epoch, but gaining on quality of photometry in the final deep tile. The final deep tile image in fact presents a constant, round and homogeneous PSF, that allows us to perform aperture or PSF photometry on the entire tile without correction on the final catalog and optimising to improve the PSF photometry. 
Usually, variations in the seeing less than $35\%$, that represent most of seeing differences in the VMC data set, are well corrected with this methodology and the degradation of the final deep tile is less than $20\%$, compared with the original (non-homogenized) deep tile.

Having a homogenized tile image is also important for the evaluation of errors and completeness through very extensive artificial star tests (AST, see Sect.~\ref{sec:psfphotometry}), preventing the occurrence of systematic errors due to differences between the PSF model used to simulate the artificial stars, and the PSF of real stars on the image. 

\begin{figure}
\resizebox{\hsize}{!}{\includegraphics{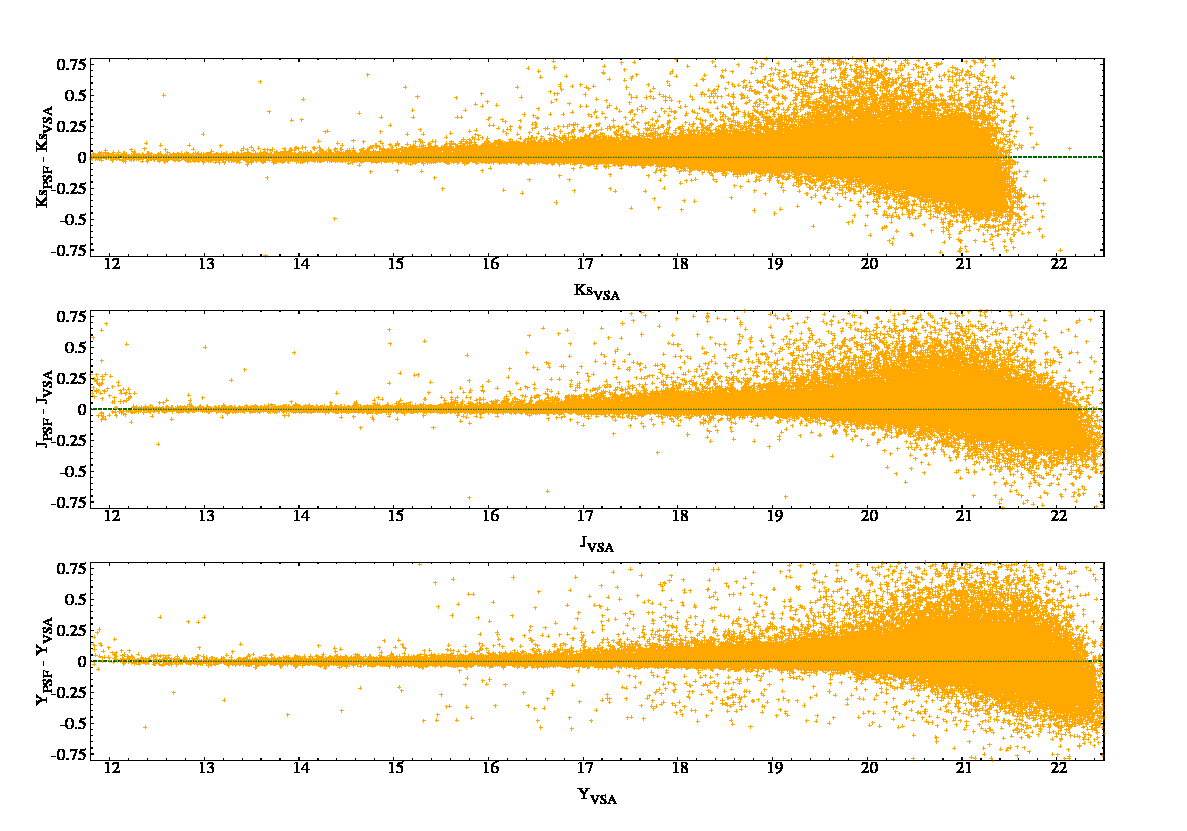}}
\caption{Difference between the PSF photometry in $Y$, $J$, and \ks\ (panels from bottom to top, respectively) and the VSA aperture photometry, for the tile SMC 3\_5, as a function of magnitude.}
\label{vsapsf}
\end{figure}

Finally, a comparison between our homogenized PSF photometry and the VSA aperture photometry is shown in Fig.~\ref{vsapsf}. It reveals that the two photometries are similar, to within 0.1~mag, down to depths of 22~mag in $Y$ and $J$, and 21.5~mag in \ks. Of course, differences are larger in high density fields or in the centres of star clusters, where the PSF photometry is expected to be more reliable.

\begin{figure*}
\resizebox{\hsize}{!}{\includegraphics{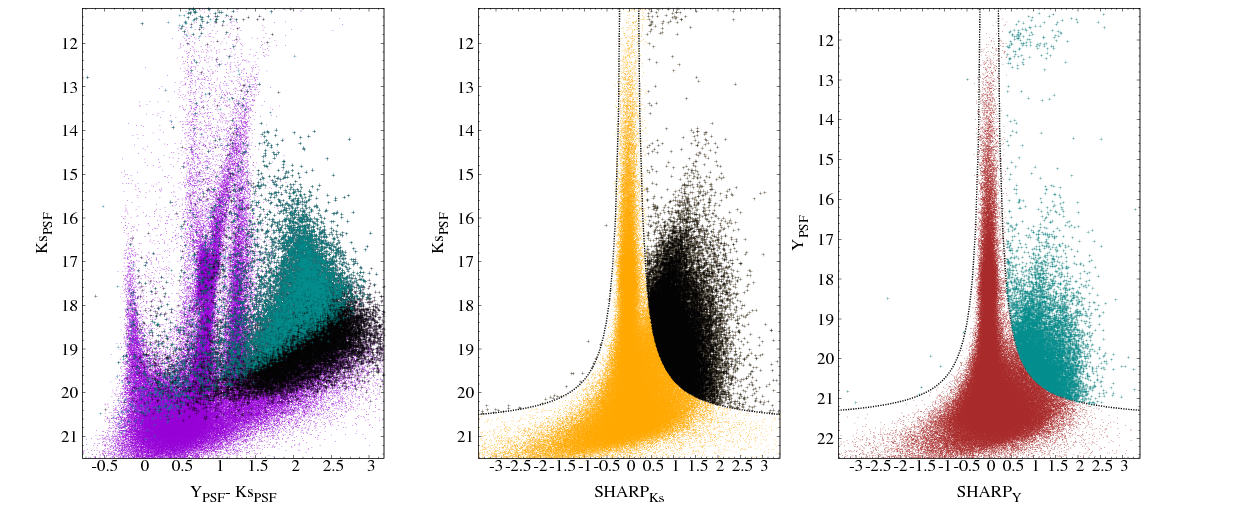}}
\caption{{\bf Left panel:} CMD of tile SMC 3\_5 showing the position of point-like objects (violet), and extended objects (black and cyan) selected using the sharp parameter in \ks\ and $Y$ bands, respectively. {\bf Central panel:} the sharpness parameter as a function of the $\ks$-band magnitude, with yellow dots marking objects selected as point-like and black dots marking the extended ones. {\bf Right panel:} the same for the $Y$ band, with burgundy and cyan dots for point-like and extended objects, respectively.}
\label{testphot}
\end{figure*}

Figure~\ref{testphot} shows an example of the efficiency of the DAOPHOT parameter ``sharp'' in separating between point-like sources and extended objects -- supposed to be stars and galaxies, respectively. Simple cuts in the sharp versus magnitude diagram (central and right panels) are able to isolate most stars, which have sharp values close to zero. Significant groups of objects with different sharpness are found both at the brightest and faintest magnitudes (black and cyan points). The former are simply partially-saturated stars at $\ks\la11.4$~mag. The latter are mainly background galaxies. Examination of the CMDs (left panel) reveals that they are mainly located in the red part of the CMD. Fortunately, the bulk of the galaxies are redder than most SMC and MW foreground stars.

\section{Photometric zeropoints}
\label{sec:zeropoints}
 
As described in \citet{Rubele_etal12}, we have estimated the differences between the calibration equations provided by CASU (eq.~\ref{eq_calib}) and those predicted by the theoretical models which are, by construction, strictly on a Vegamag system (where Vega star has a null magnitude in all filters). The same process is repeated for this work, since we are using a different version of VSA data, and improved photometry. The basic idea is to reproduce, with the aid of stellar models, the same sort of photometric data used by CASU in the original calibration of VISTA data from 2MASS observations.

\begin{figure}
\resizebox{\hsize}{!}{\includegraphics{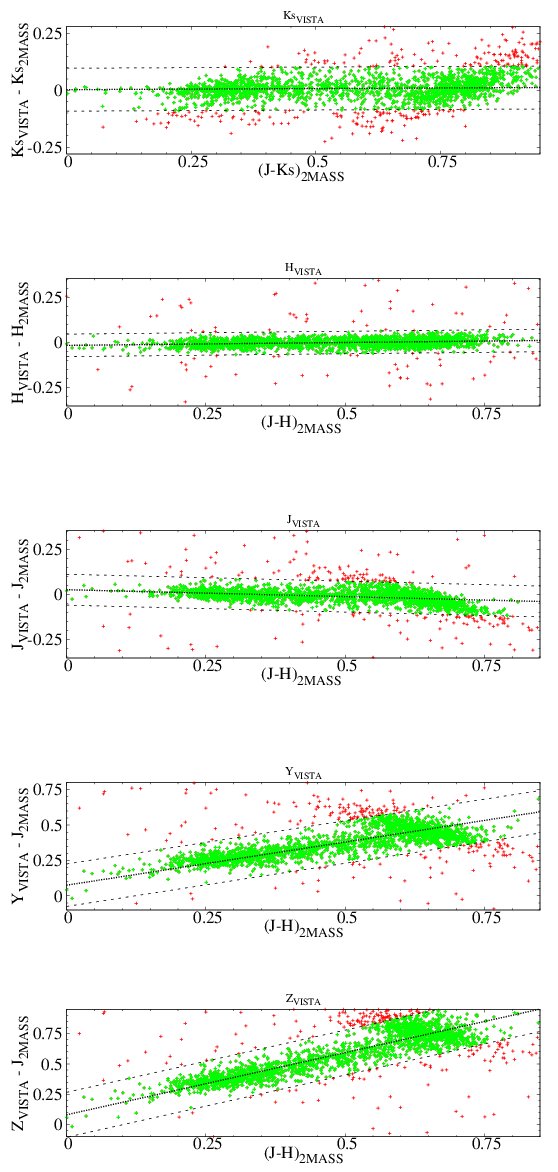}}
\caption{Calibration to a Vegamag system. The points show the difference between VISTA and 2MASS magnitudes versus the 2MASS colours of simulated Milky Way model stars on an area of $\sim 1$~\sqdeg. The simulated stars are generated assuming the observed distrbution of 2MASS errors. Green points show the distribution of stars inside the adopted $3\sigma$ clipping limit (black dotted lines). The black dashed dotted line shows the best-fit linear relation, with slopes fixed as in Eqs.~\ref{eq_calib}.}
\label{testcalib}
\end{figure}

We proceed in the following way: First we generate a synthetic model of the Milky Way (MW), containing both 2MASS and VISTA filters, using the TRILEGAL code \citep{Girardi_etal05}. The simulated region has a total area of $\sim 1.0$~\sqdeg\ towards the SMC. The typical 2MASS photometric errors were added to the simulations following \cite{bon04}. They dominate the error distribution compared to VISTA data. Then, we fit the distributions of these points in the colour-colour diagrams shown in Fig.~\ref{testcalib}, keeping the slope fixed at the same value as adopted by CASU, and adopting a sigma clipping of $3\sigma$. For the best-fit equations we obtain:
\begin{eqnarray}
 Z_{\rm VISTA}-J_{\rm 2MASS} & = & 1.025(J\!-\!H)_{\rm 2MASS} + 0.082 \nonumber\\
 Y_{\rm VISTA}-J_{\rm 2MASS} & = & 0.610(J\!-\!H)_{\rm 2MASS} + 0.074 \label{eq_calib} \\
 J_{\rm VISTA}-J_{\rm 2MASS} & = & -0.077(J\!-\!H)_{\rm 2MASS} + 0.026 \nonumber\\
 H_{\rm VISTA}-H_{\rm 2MASS} & = & 0.032(J\!-\!H)_{\rm 2MASS} + -0.017 \nonumber \\
 \ks_{\rm VISTA}-\ks_{\rm 2MASS} & = & 0.010(J\!-\!\ks)_{\rm 2MASS} + 0.003 \nonumber
\end{eqnarray}

These fits imply offsets of $0.082$ mag in $Z$, $0.074$ in $Y$, $0.026$ in $J$, $-0.017$ in $H$ and $0.003$ in \ks, between the model Vega-magnitudes and the CASU calibrations, reflecting the fact that the underlying colour-colour relations are not strictly linear as assumed in the calibrating equations. We apply these offsets to our stellar models (Sect.~\ref{sec:partialmodels}), thus converting them to the same set of ZPs as the observations, before starting the work of SFH-recovery.\footnote{Although we just use the offsets in $Y$, $J$ and $\ks$, those in $Z$ and $H$ are also provided here, so that everybody can convert the PARSEC isochrones (\url{http://stev.oapd.inaf.it/cmd}) -- which are provided in a strict Vegamag VISTA system -- into something more similar to the v1.1 and v1.2 VSA data.}
 
%
%

%
\label{lastpage}
\end{document}